\documentclass[]{pasj02} 
\usepackage[switch,mathlines]{lineno} 
\usepackage{natbib} 
\usepackage{graphicx}
\usepackage{amsmath}
\usepackage{bm}
\usepackage{color}

\defcitealias{Linial&Metzger2023}{LM23}
\defcitealias{Strubbe&Quataert2009}{SQ09}

\jyear{2025}
\Received{}
\Accepted{}

\begin{document} 

\title{Possibility of Multi-Messenger Observations of Quasi-Periodic Eruptions with X-rays and Gravitational Waves}

\author{
 Tomoya \textsc{Suzuguchi},\altaffilmark{1}\altemailmark\orcid{0009-0005-1459-1846} \email{suzuguchi@tap.scphys.kyoto-u.ac.jp}
 Hidetoshi \textsc{Omiya},\altaffilmark{1}\orcid{0000-0003-2437-3621}
 and
 Hiroki \textsc{Takeda},\altaffilmark{1,2}\orcid{0000-0001-9937-2557}
}
\altaffiltext{1}{Department of Physics, Kyoto University, Sakyo, Kyoto 606-8501, Japan}
\altaffiltext{2}{The Hakubi Center for Advanced Research, Kyoto University, Sakyo, Kyoto 606-8501, Japan}




\maketitle

\begin{abstract}
Recent X-ray observations have discovered a class of periodic X-ray flares in galactic nuclei known as quasi-periodic eruptions (QPEs). A promising explanation of QPEs is an emission produced when a stellar-mass object orbiting a central supermassive black hole crosses an accretion disk. If the companion is a compact object, such systems would be a prospective multi-messenger target for the space-based observatory LISA and its successors. Here we quantify the prospects for joint X-ray and GW detection of QPEs with orbital frequency in the mHz band using a minimal flare-emission model. Our analysis shows that X-ray observations are most effective at orbital frequencies up to roughly 1 mHz, whereas LISA is sensitive chiefly above about 1 mHz. Because the optimal sensitivity windows overlap only marginally, we predict at most one joint detection during LISA’s nominal mission lifetime. Extending GW sensitivity into the sub-millihertz regime (< 0.1 mHz) would raise the possibility of the joint detection by an order of magnitude, enabling QPEs as an interesting multi-messenger target. 
\end{abstract}


\section{Introduction} \label{sec:intro}
Quasi-periodic eruptions (QPEs) are recurrent X-ray transients originating from galactic nuclei, characterized by periodic bursts in their X-ray light curves occurring every few hours to several days. Since their first discovery by \textit{XMM-Newton} in December 2018~\citep{Miniutti+2019}, nearly a dozen confirmed or candidate QPEs have been detected by \textit{XMM-Newton}, \textit{Chandra}, \textit{NICER}, and \textit{eROSITA}~\citep{Giustini+2020, Chakraborty+2021, Arcodia+2021, Quintin+2023, Arcodia+2024a, Nicholl+2024, Bykov+2024, Chakraborty+2025, Hernandez-Garcia+2025}. These events typically exhibit peak X-ray luminosities of $10^{41}-10^{43}\,{\rm erg}\,{\rm s}^{-1}$ and quasi-thermal spectra with characteristic temperatures in the range $100-200\,{\rm eV}$. The duty cycle of QPEs has been estimated to be $\sim 10\%$, a value that appears consistent across different sources. Based on blind searches by \textit{eROSITA}, their volumetric formation rate is estimated to be $\sim 6\times10^{-8}\,{\rm Mpc}^{-3}\,{\rm yr}^{-1}$, assuming each QPE persists for approximately $10$ years~\citep{Arcodia+2024b}. 

The physical origin of QPEs remains uncertain, with several theoretical models having been proposed. These include limit-cycle oscillations driven by disk instabilities~\citep{Raj&Nixon2021,Pan+2022,Pan+2023,Kaur+2023,Sniegowska+2023}, gravitational self-lensing effects caused by supermassive black holes (SMBHs)~\citep{Ingram+2021}, Lense-Thirring precession of super-Eddington outflows~\citep{Middleton+2025}, and episodic mass transfer from evolved stars onto SMBHs~\citep{King2020,Krolik&Linial2022,Metzger+2022,Linial&Sari2023,Lu&Quataert2023,Olejak+2025}. Among the proposed explanations, a particularly compelling scenario suggests that QPEs result from interactions between a star orbiting a SMBH and the accretion disk~\citep{Dai+2010,Xian+2021,Linial&Metzger2023,Franchini+2023,Tagawa&Haiman2023,Linial&Metzger2024b,Zhou+2024a,Zhou+2024b,Zhou+2024c,Vurm+2025,Yao+2025}. In this ``EMRI+disk'' scenario---where EMRI refers to an extreme mass ratio inspiral---X-ray flares are triggered each time the inspiraling star (hereafter simply ``the EMRI'') periodically passes through the disk. This model has shown promising consistency with the long-term behavior of observed QPEs~\citep[e.g.,][]{Chakraborty+2024,Arcodia+2024c}, suggesting that the EMRI+disk model captures essential aspects of the underlying physics. 

Recent observations have revealed a potential connection between QPEs and tidal disruption events (TDEs). Some QPEs have been discovered in galaxies that previously hosted TDEs, with time intervals of a few years between the two phenomena~\citep{Miniutti+2019,Chakraborty+2021,Quintin+2023,Nicholl+2024,Bykov+2024,Chakraborty+2025}. Moreover, QPE host galaxies exhibit a strong preference for low-mass, post-starburst/quiescent Balmer-strong galaxies---mirroring the demographics of known TDE hosts~\citep{Wevers+2022, Wevers+2024}. These correlations suggest that the accretion disks involved in the EMRI+disk model may originate from TDEs. Given the wide range of physical parameters associated with TDE disks, including time-dependent accretion rates, QPEs may emit across a broad wavelength range, potentially extending from the UV to the hard X-ray bands. While QPEs have thus far been detected primarily in the soft X-ray regime, their broader spectral signatures may become accessible with future observations~\citep{Linial&Metzger2024a, Suzuguchi&Matsumoto2025}. 

Separately, EMRIs are recognized as promising sources for future space-based gravitational wave (GW) observatories such as the \textit{Laser Interferometer Space Antenna} (LISA)~\citep[e.g.,][]{Amaro-Seoane+2017,Amaro-Seoane+2023}. Combined with the growing evidence linking QPEs to EMRI activity, this raises the exciting possibility that QPEs could serve as multi-messenger sources. If QPEs are detected in both X-rays and GWs, a wealth of information could be obtained. For example, GW signals provide robust measurements of the SMBH masses, which can also be independently inferred from QPE timing properties~\citep{Zhou+Pan+Jiang2025}. Furthermore, multi-messenger detections could offer valuable cosmological insights, such as measurements of the Hubble constant~\citep{Lyu+2025}. 

Although multi-messenger observations of QPEs have the potential to yield valuable insights, it remains uncertain whether such events can be jointly detected by future GW and X-ray observatories. A key challenge lies in the fact that recently discovered QPEs exhibit orbital frequencies in the microhertz ($\mu$Hz) regime, while LISA is sensitive to signals in the millihertz (mHz) band ($10^{-1}$–$10$ mHz). To address this mismatch, we focus on QPEs with recurrence timescales of approximately $0.1\,{\rm hrs}$, which we refer to as \textit{mHz-QPEs}, in contrast to the previously known events, hereafter denoted as \textit{$\mu$Hz-QPEs}. A recent study by~\citet{Kejriwal+2024} also considered the multi-messenger study of QPEs. However, their analysis was limited to assessing whether EMRIs observable by LISA in the late 2030s could correspond to systems currently manifesting as $\mu$Hz-QPEs, and did not consider the feasibility of simultaneous GW and X-ray detection.

In this work, we investigate the prospects for the concurrent detection of mHz-QPEs in both gravitational waves and X-rays, within the framework of the EMRI+disk model. We primarily consider systems in which the EMRI consists of a stellar-mass black hole, as these are the most promising sources of GW emission. Employing the emission model developed in~\citet{Linial&Metzger2023}, we estimate the expected detection rate of mHz-QPEs. Our results indicate that although mHz-QPEs may be detectable by future X-ray observatories, the likelihood of multi-messenger detections is limited, with a number of expected events to be $\mathcal{O}(10^{-1})$ within the nominal mission lifetime of LISA. Moreover, we highlight that the successful multi-messenger detection of a QPE could place stringent constraints on the propagation speed of gravitational waves in the mHz frequency band.

The organization of this paper is as follows. In Section~\ref{sec:emission}, we briefly review the emission mechanism underlying the EMRI+disk model and discuss the X-ray spectrum of QPEs. In Section~\ref{sec:observability}, we assess the multi-messenger detectability of QPEs at millihertz GW frequencies. We also discuss the possibility of constraining the propagation speed of GWs in the millihertz band as an example of the scientific opportunities enabled by multi-messenger detections of QPEs in this section. In Section~\ref{sec:other}, we extend the discussion to alternative scenarios, including cases where the EMRI is a star, white dwarf (WD), or neutron star (NS), and where the accretion disk originates from active galactic nuclei (AGN) rather than TDEs. Finally, in Section~\ref{sec:conclusion}, we conclude with a summary and discussion.

\section{Emission Model}  \label{sec:emission}
\begin{figure*}[t]
  \begin{center}
  \includegraphics[keepaspectratio, scale=0.3]{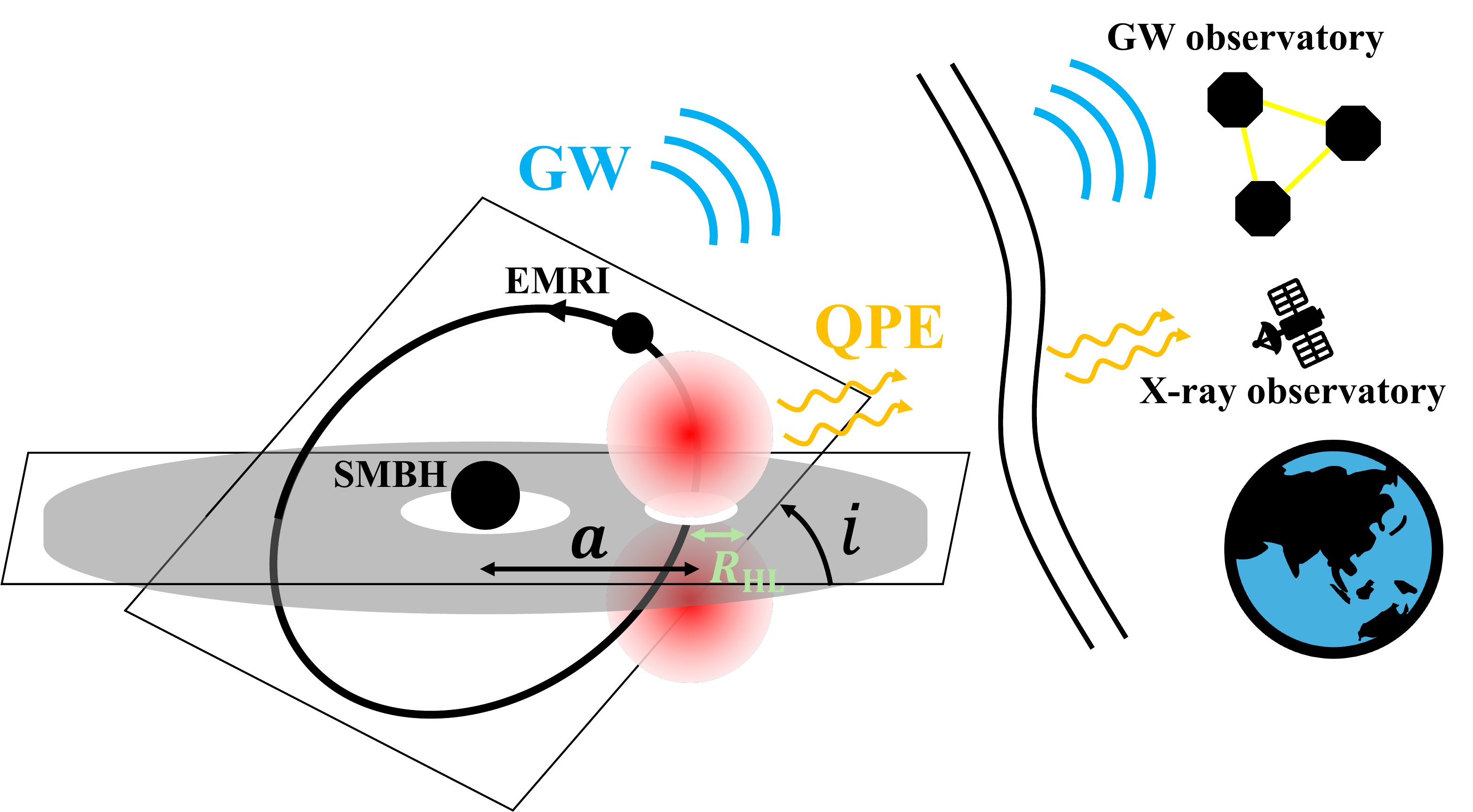}
  \end{center}
  \caption{Schematic illustration of the system considered in this work. A stellar-mass BH (simply referred to as an EMRI) orbits a central SMBH and collides with the accretion disk twice per orbit, ejecting disk gas within its Hoyle-Lyttleton radius. As the initially opaque ejecta expands and becomes less dense, photons are emitted, producing the observed X-ray QPEs. Simultaneously, the SMBH-EMRI system emits GWs, making it a potential multi-messenger source for future space-based GW observatories. This study examines the feasibility of simultaneous detection of GWs and X-rays from such systems. }
  \label{fig:schematic}
\end{figure*}

In this section, we estimate the duration, luminosity, and temperature of QPEs based on the ``EMRI+disk'' model proposed by~\citet{Linial&Metzger2023} (hereafter~\citetalias{Linial&Metzger2023}). In this framework, QPEs are triggered by collisions between a star orbiting a central SMBH and the accretion disk. As the star traverses the disk at supersonic velocities, shock waves are generated, causing significant compression of the disk gas. This compression drives a rapid expansion of the gas both above and below the disk. Initially, the expanding gas remains optically thick, trapping photons and preventing their escape. As the gas becomes less opaque, photons are released, producing the observable QPEs. The situation is described in Figure~\ref{fig:schematic}. 

We consider a BH companion with mass $M_{\star}=10M_{\odot}m_{\star,1}$, in contrast to the solar-like stellar companion assumed in \citetalias{Linial&Metzger2023}. Here $M_{\odot}$ is the solar mass. In the following, we refer to this BH companion as an EMRI. The EMRI orbits a central SMBH with mass $M_{\bullet}=10^6M_{\odot}M_{\bullet,6}$. For simplicity, we take the central SMBH to be non-spinning. The GW frequency of the EMRI is 
\begin{align}
    \label{eq:f_gw}
    f_{\rm GW} &= \frac{1}{\pi}\left(\frac{GM_{\bullet}}{a^3}\right)^{1/2} \nonumber \\
    &\simeq 2.0\,{\rm mHz}\,
    \frac{1}{a_{1}^{3/2} M_{\bullet,6}},
\end{align}
\citep[e.g.,][]{Maggiore2007} where $a=10R_{\rm g}a_{1}$ is the semimajor axis of the EMRI orbit, and $R_{\rm g}=GM_{\bullet}/c^2$ is the gravitational radius. Here, $G$ and $c$ are the gravitational constant and the speed of light, respectively. The inclination angle between the EMRI's orbital plane and the disk midplane is denoted by $i$. 

The accretion disk is assumed to be a standard disk~\citep{Shakura&Sunyaev1973}, which, as discussed in Section~\ref{subsec:parameter}, is expected to form a few years after a TDE. If the EMRI remains within the LISA band, the disk region at the crossing point is dominated by radiation pressure. In this case, the aspect ratio and surface density of the disk are 
\begin{align}
    \label{eq:scale_height}
    \frac{H}{R} &\simeq \frac{3}{2\epsilon}\frac{R_{\rm g}}{R}\frac{\dot{M}}{\dot{M}_{\rm Edd}}g \simeq 0.15\frac{\dot{m}_{-1}g}{a_{1}}, \\
    \label{eq:surface_density}
    \Sigma &= \frac{\dot{M}}{3\pi\nu}g 
    \simeq 5.5\times10^{2}\,{\rm g}\,{\rm cm}^{-2}
    \frac{a_{1}^{3/2}}{\alpha_{-1}\dot{m}_{-1}g},
\end{align}
where $H$ is the vertical scale height of the disk, $R$ is the radial distance from the central SMBH, $g=1-(6R_{\rm g}/R)^{1/2}$, $\dot{m}=\dot{M}/\dot{M}_{\rm Edd}$ is the accretion rate normalized by the Eddington rate, $\dot{M}_{\rm Edd}=L_{\rm Edd}/(\epsilon c^{2})$, and $\nu=\alpha c_{\rm s}H$ is the kinematic viscosity of the disk, where $c_{\rm s}=H\Omega_{\rm K}$ is the sound speed of the disk, $\alpha$ is the $\alpha$-parameter~\citep{Shakura&Sunyaev1973}, and $\Omega_{\rm K}=(GM_{\bullet}/a^{3})^{1/2}$ is the Kepler angular velocity. Here, $L_{\rm Edd}=4\pi cGM_{\bullet}/\kappa_{\rm es}$ is the Eddington luminosity where $\kappa_{\rm es}\simeq 0.34\,{\rm cm}^{2}\,{\rm g}^{-1}$ is the electron scattering opacity and $\epsilon=0.1$ is the radiative efficiency. 

The relative velocity between the EMRI and the disk during a disk crossing is $v_{\rm rel}=2\sin{(i/2)}v_{\rm K}$, where $v_{\rm K}=(GM_{\bullet}/a)^{1/2}$ is the Keplerian velocity. As the BH traverses the disk, it shocks the surrounding disk material with mass $M_{\rm ej}=2\pi(R_{\rm inf}/\sin{i})^2\Sigma$, where $R_{\rm inf}$ is the radius of influence. If the EMRI is a BH, the radius of influence is given by the Hoyle-Lyttleton (HL) radius, defined as 
\begin{align}
    \label{eq:HL_radius}
    R_{{\rm HL},\star} &= \frac{2GM_{\star}}{v_{\rm rel}^{2}} \nonumber \\
    &\simeq 2.1\times10^{-4}R_{\odot}\,
    \frac{a_{1}m_{\star,1}}{\sin^{2}{(i/2)}},
\end{align}
\citep{Hoyle&Lyttleton1939, Bondi&Hoyle1944}. The shocked material, initially optically thick, expands vertically from the disk and reaches a terminal velocity $v_{\rm ej}\simeq v_{\rm rel}$. Photons begin to escape when the optical depth, which scales as $\tau\propto t^{-2}$, drops to $\tau\simeq c/v_{\rm ej}$. The corresponding timescale is 
\begin{align}
    \label{eq:duration}
    t_{\rm QPE} &\simeq \left(\frac{\kappa_{\rm es} M_{\rm ej}}{4\pi cv_{\rm ej}}\right)^{1/2} \nonumber \\
    &\simeq 2.7\times10^{-6}\,{\rm hrs}\,
    \frac{1}{\sin{i}\sin^{5/2}{(i/2)}} \nonumber \\
    &\quad\times\frac{ \left(R_{\rm inf}/R_{{\rm HL},\star}\right) m_{\star,1} a_{1}^{2}}{\alpha_{-1}^{1/2}\dot{m}_{-1}^{1/2}g^{1/2}},
\end{align}
which characterizes the duration of QPEs. The QPE luminosity is estimated from the internal energy of the shocked material, accounting for adiabatic losses. Considering that the initial volume of the shocked gas is ${\cal V}_{0}\simeq\pi(R_{\rm inf}/\sin{i})^{2}H/7$, and that the pressure is radiation-dominated, the internal energy at the time of emission is given by $E_{\rm QPE}\simeq({\cal V}_{0}/(4\pi(v_{\rm ej}t_{\rm QPE})^{3}))^{1/3}E_{\rm ej}$, where the initial internal energy is $E_{\rm ej}\simeq (1/2)M_{\rm ej}v_{\rm ej}^{2}$. The corresponding QPE luminosity is therefore 
\begin{align}
    \label{eq:luminosity}
    L_{\rm QPE} & \simeq \frac{E_{\rm QPE}}{t_{\rm QPE}} \nonumber \\
    & \simeq 2.2\times10^{40}\,{\rm erg}\,{\rm s}^{-1}\,
    \frac{\sin^{2/3}{(i/2)}}{\sin^{2/3}{i}} \nonumber \\
    &\quad\times\frac{M_{\bullet,6}^{1/3}\left(R_{\rm inf}/R_{{\rm HL},\star}\right)^{2/3}m_{\star,1}^{2/3}\dot{m}_{-1}^{1/3}g^{1/3}}{a_{1}^{1/3}}.
\end{align}

The blackbody temperature is given by 
\begin{align}
    T_{\mathrm{BB}} &= \left(\frac{u_{\gamma}}{a_{\rm rad}}\right)^{1/4} \nonumber \\
    &\simeq 4.9\times10^{2}\,{\rm eV}\,
    \sin^{1/3}{i}\sin^{2/3}{(i/2)} \nonumber \\
    &\quad\times\frac{\alpha_{-1}^{1/4}
    M_{\bullet,6}^{1/12}\dot{m}_{-1}^{1/3}g^{1/3}}{\left(R_{\rm inf}/R_{{\rm HL},\star}\right)^{1/3} m_{\star,1}^{1/3}a_{1}^{17/24}}.
\end{align}
where $u_{\gamma}=L_{\rm QPE}/[4\pi(v_{\rm ej}t_{\rm QPE})^2v_{\rm ej}]$ is the radiation energy density at $t\simeq t_{\rm QPE}$, and $a_{\rm rad}$ is the radiation constant. If photon production is sufficiently efficient to establish thermal equilibrium, the observed temperature coincides with the blackbody temperature. However, if the photon production is inefficient, thermalization is incomplete and the radiation temperature can exceed the blackbody value. This can be quantified using the so-called ``thermalization efficiency''~\citet{Nakar&Sari2010}:
\begin{align}
    \label{eq:eta}
    \eta &= \frac{n_{\rm BB}(T_{\rm BB,sh})}{t_{\rm cross}\dot{n}_{\gamma,{\rm ff}}(\rho_{\rm sh},T_{\rm BB,sh})} \nonumber \\
    &\simeq 3.7\times10^2\,
    \sin{i}
    \frac{\alpha_{-1}^{9/8} M_{\bullet,6}^{1/8} \dot{m}_{-1}^{5/4}g^{5/4}}{a_{1}^{49/16}},
\end{align}
which is the ratio of the photon number density required for thermal equilibrium, $n_{\rm BB}(T_{\rm BB})=a_{\rm rad}T_{\rm BB}^{4}/(3k_{\rm B}T_{\rm BB})$, to the photon number density produced by the free-free emission over the expansion timescale, $t_{\rm cross}\simeq (H/7)/(v_{\rm K}\sin{i})$. Here, $k_{\rm B}$ is the Boltzmann constant, and $T_{\rm BB,sh}\simeq(3\rho v_{\rm K}^{2}/a_{\rm rad})^{1/4}$ and $\rho_{\rm sh}\simeq 7\rho$ are the blackbody temperature and density of the ejecta immediately after the shock passage, with the pre-shock gas density of $\rho\simeq \Sigma/(2H)$. In the second line, we adopt the free-free photon production rate of $\dot{n}_{\gamma,{\rm ff}}(\rho,T)\simeq 3.5\times10^{36}\,{\rm s}^{-1}\,{\rm cm}^{-3}\,\rho^2T^{-1/2}$~\citep[see][]{Nakar&Sari2010}. 

The thermalization efficiency $\eta$ exceeds unity for the fiducial parameters considered here, indicating inefficient photon production and a deviation from thermal equilibrium. In this case, the observed temperature is given by 
\begin{align}
    \label{eq:obs_temperature}
    T_{\rm obs} &\simeq \eta^2T_{\rm BB} \nonumber \\
    &\simeq 8.0\,{\rm MeV}\,
    \sin^{7/3}{i}\sin^{2/3}{(i/2)}
    \frac{\alpha_{-1}^{5/2} M_{\bullet,6}^{1/3} \dot{m}_{-1}^{17/6}g^{17/6}}{\left(R_{\rm inf}/R_{{\rm HL},\star}\right)^{1/3}m_{\star,1}^{1/3}a_{1}^{41/6}}.
\end{align}
However, this prescription is applicable only when $T_{\rm obs}\lesssim 50\,{\rm keV}$~\citep{Nakar&Sari2010}. For higher temperatures, pair production becomes significant, suppressing any further increase in temperature~(e.g., \citealt{Katz+2010,Budnik+2010,Levinson&Nakar2020} for a review). In such cases, the temperature saturates at $\simeq m_{\rm e}c^{2}/3$, where $m_{\rm e}$ is the electron mass~\citep[e.g.,][]{Ito+2018,Ito+2020}. In summary, the observed QPE temperature can be expressed as 
\begin{equation}
\label{eq:QPEtemp}
    \begin{split}
        k_{\rm B}T_{\rm QPE} = 
        \begin{cases}
            {\rm max}(k_{\rm B}T_{\rm obs}, k_{\rm B}T_{\rm BB}), & {\rm for}\,\,\,T_{\rm obs}<50\,{\rm keV}, \\
            m_{\rm e}c^{2}/3, & {\rm for}\,\,\,T_{\rm obs}\gtrsim 50\,{\rm keV}.
        \end{cases}
    \end{split}
\end{equation}

\subsection{Parameter dependencies of QPE observables} \label{subsec:parameter}
In our framework, the QPE observables $t_{\rm QPE}, L_{\rm QPE}$, and $T_{\rm QPE}$, are characterized by the SMBH mass $M_{\bullet}$, EMRI mass $M_{\star}$, the orbital separation $a$ between the SMBH and the EMRI, viscosity parameter $\alpha$, inclination angle $i$, and accretion rate $\dot{M}$. In this study, we fix the mass of the SMBH to $M_{\bullet}=10^{6}M_{\odot}$, consistent with typical values inferred for $\mu$Hz-QPE host galaxies. The EMRI mass, by contrast, is more uncertain. However, various estimates of the stellar-mass BH mass function suggest a peak in the range $10-30 M_{\odot}$~\citep[e.g.,][]{Sicilia+2022}. We therefore adopt a fiducial EMRI mass of $M_{\star}=10M_{\odot}$. Regarding the orbital separation, our focus is on EMRIs detectable by LISA, which is sensitive to GW frequencies in the range $0.1-10\,{\rm mHz}$. For a fixed SMBH mass, this frequency range corresponds to orbital separations of $a\simeq (3.5-75)R_{\rm g}$. In what follows, we consider two representative cases: $a=10R_{\rm g}$ (corresponding to $f_{\rm GW}=2\,{\rm mHz}$, the sweet spot of LISA's sensitivity) and $a=50R_{\rm g}$ (corresponding to $f_{\rm GW}=0.2{\rm mHz}$, near the lower end of LISA's sensitivity). The $\alpha$-parameter is also poorly constrained, but we follow \citetalias{Linial&Metzger2023} in adopting $\alpha=0.1$.

The remaining two parameters, the inclination angle $i$ and the accretion rate $\dot{M}$, are treated as free parameters. As far as the authors are aware, there is no preferred alignment between the orbital plane of the EMRI and the angular momentum of the disk. Thus, the inclination angle can be assumed to be randomly distributed over the full range $0^\circ\leq i\leq 180^\circ$, where $0^\circ\leq i\leq 90^\circ$ and $90^\circ \leq i \leq 180^\circ$ correspond to prograde and retrograde orbits relative to the disk's  rotation, respectively. However, if the inclination is smaller than the opening angle of the disk, $\tan^{-1}{(H/R)}\simeq 8.5^{\circ}$, the EMRI would be completely embedded within the disk, and no quasi-periodic signals would be produced. Since such systems are beyond the scope of this paper, we restrict our focus to the systems with inclination angles in the range $10^{\circ}\leq i \leq 170^{\circ}$. 

At each disk crossing, the inclination of the EMRI may change due to dynamical friction exerted by the disk gas. The timescale for the evolution of the inclination can be estimated using the dynamical friction formula~\citep[e.g.,][]{Ostriker1999,Kim&Kim2009,Thun+2016,Suzuguchi+2024}. Previous studies employing such estimates~\citep[e.g.,][]{Subr&Karas1999,Karas&Subr2001, Generozov&Perets2023, Spieksma&Cannizzaro2025} have shown that the inclination change is generally slower than the orbital evolution driven by GW emission, at least as long as the EMRI remains within the LISA band. This is because the dynamical friction acting on the EMRI in the LISA band is weaker due to higher velocities and lower gas densities (see Equations.~\eqref{eq:scale_height} and \eqref{eq:surface_density}). Therefore, as long as the initial inclination exceeds the disk opening angle, subsequent embedding into the disk is unlikely and can be safely neglected in our analysis. 

Moreover, the accretion rate of a TDE disk generally declines over time, meaning that the accretion rate evolves from high to low values as time elapses. Immediately after the formation of the disk, the accretion rate can exceed the Eddington limit ($\dot{m}>1$) \citep[e.g.,][]{Rees1988,Phinney1989,Evans&Kochanek1989}, resulting in a geometrically thick, radiation-pressure-dominated disk~\citep[e.g.,][]{Abramowicz+1988}. After a few years, as the accretion rate falls below the Eddington limit, the disk transitions into a standard, geometrically thin, radiatively efficient state~\citep[e.g.,][]{Strubbe&Quataert2009,Shen&Matzner2014}. If the accretion rate continues to decline further, the disk can eventually enter a radiatively inefficient accretion flow (RIAF) regime, characterized by a higher midplane temperature due to inefficient radiative cooling~\citep[e.g., see][]{Hayasaki&Yamazaki2019}. This transition is expected to occur around a critical accretion rate of $\dot{m}_{\rm crit}\sim\alpha^2=0.01\alpha_{-1}^{2}$~\citep[e.g.,][]{Narayan&Yi1994,Narayan&Yi1995}. Given that we focus on the standard disk phase in this work, we restrict our analysis to the accretion rate range $10^{-3}\leq\dot{m}\leq1$, while accounting for potential uncertainties of up to a factor of a few. 

This choice is motivated by the following considerations. In the super-Eddington regime, the QPE temperature is expected to reach the hard X-ray band~\citep{Suzuguchi&Matsumoto2025}, where detection is challenging due to the relatively low sensitivity of current hard X-ray and soft gamma-ray observatories. We therefore exclude the super-Eddington phase. Conversely, in the RIAF regime, the high sound velocity resulting from the elevated midplane temperature can inhibit shock formation as the EMRI crosses the disk. Since the shock formation is essential for generating QPE signals within our framework, we exclude the RIAF phase from our analysis. 

\begin{figure}[t]
  \begin{center}
  \includegraphics[keepaspectratio, scale=0.35]{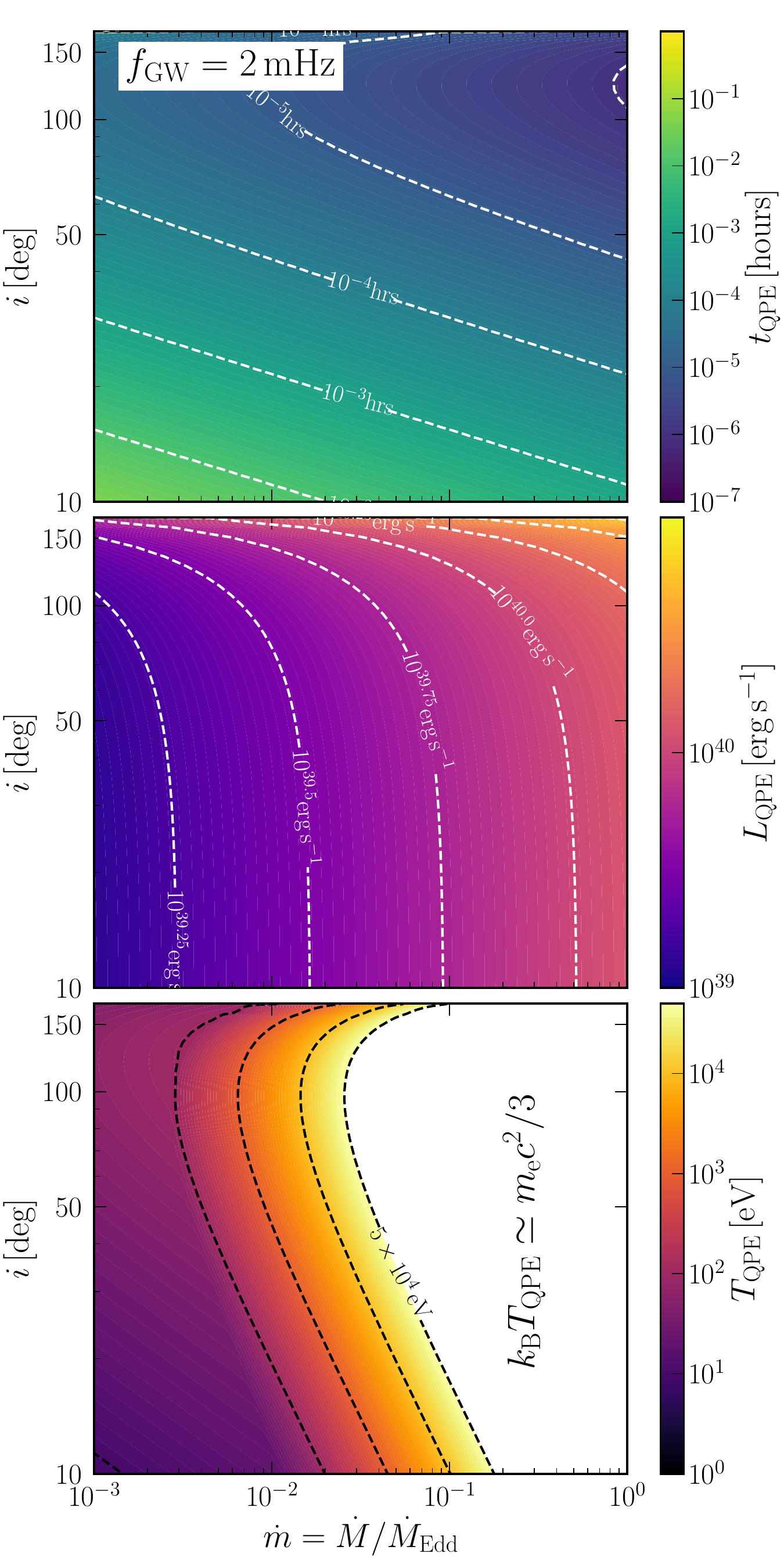}
  \end{center}
  \caption{Dependence of QPE observables on accretion rate and inclination: duration (top panel), luminosity (middle panel), and temperature (bottom panel) for $f_{\rm GW}=2\,{\rm mHz}$ (corresponding to semimajor axis $a = 10 R_{\rm g}$). As noted in the text, the other parameters of the QPE model are fixed at $M_{\bullet}=10^6M_{\odot}$, $M_{\star}=10M_{\odot}$, and  $\alpha = 0.1$. }
  \label{fig:color_2mHz}
\end{figure}

\begin{figure}[t]
  \begin{center}
  \includegraphics[keepaspectratio, scale=0.35]{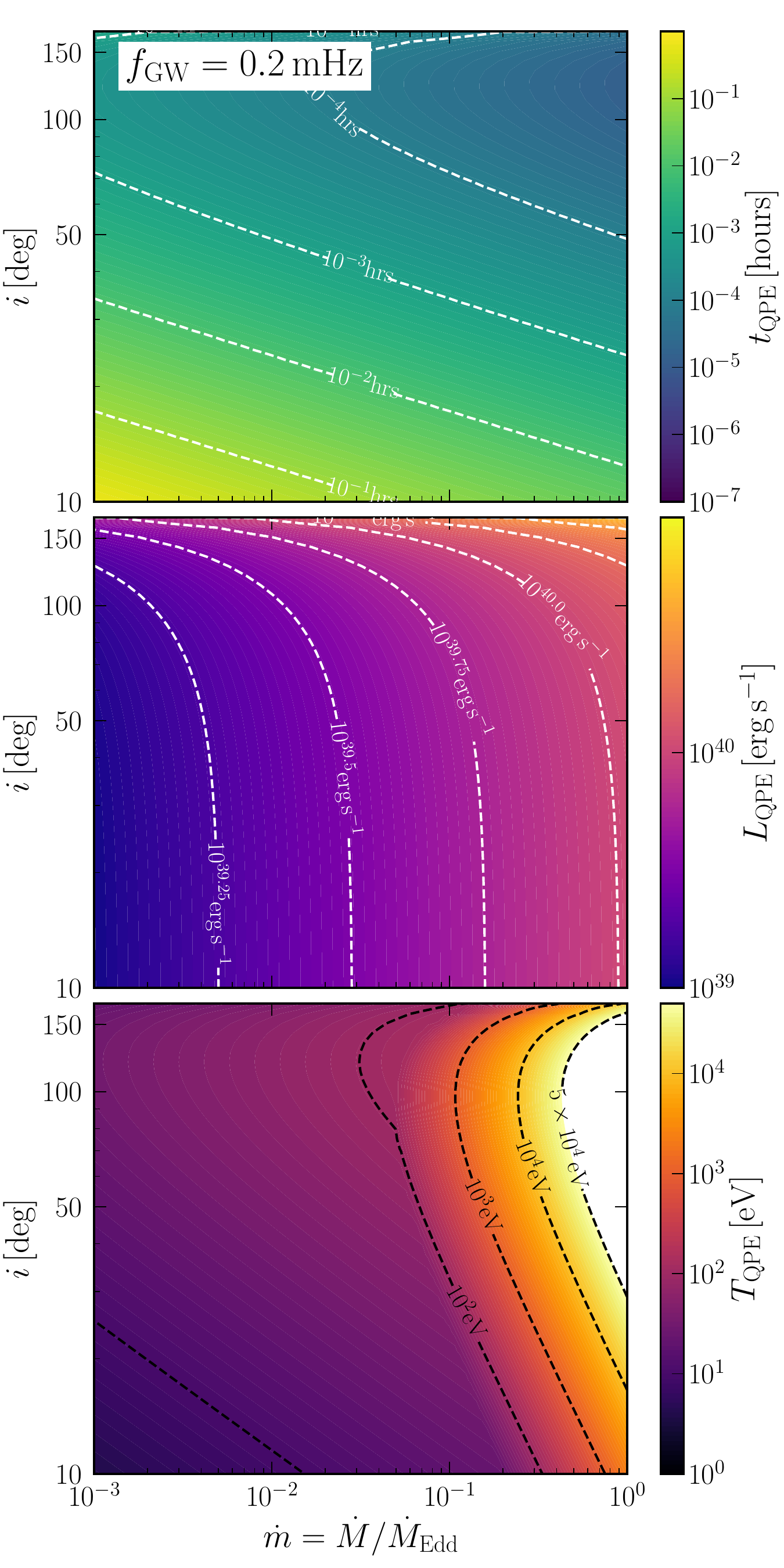}
  \end{center}
  \caption{Same as Figure~\ref{fig:color_2mHz}, except that the separation is changed such that the GW frequency is $f_{\rm GW}=0.2\,{\rm mHz}$ (corresponding to semimajor axis $a = 50 R_{\rm g}$). }
  \label{fig:color_0.2mHz}
\end{figure}

Figures~\ref{fig:color_2mHz} and \ref{fig:color_0.2mHz} show the dependence of the QPE observables on the model parameters for two representative cases: $a=10R_{\mathrm{g}}$ (corresponding to $f_{\mathrm{GW}}\simeq 2\,\mathrm{mHz}$) and $a=50R_{\mathrm{g}}$ (corresponding to $f_{\mathrm{GW}}\simeq 0.2\,\mathrm{mHz}$). The QPE duration increases as the inclination angle decreases. This is primarily due to two factors: the reduced velocity of the EMRI and the resulting increase in ejecta mass. In systems with a slightly inclined prograde orbit, the relative velocity between the EMRI and the disk is lower, which leads to slower ejecta. Such slower ejecta expand more gradually, delaying photon diffusion and prolonging the time required for the material to become optically thin. Consequently, the QPE duration is extended. Additionally, the lower velocity of the EMRI increases the radius of influence (equal to the HL radius), which allows a larger volume of disk material to be shocked and ejected. As a result, the ejecta mass becomes larger. Therefore, both the slower expansion and the increased ejecta mass contribute to the longer QPE durations observed in systems with smaller inclination angles. 

On the other hand, the QPE luminosity $L_{\rm QPE}$ is almost constant with respect to the inclination (slightly increasing with larger inclination), but shows a mild dependence on the accretion rate (see middle panel of Figures~\ref{fig:color_2mHz} and~\ref{fig:color_0.2mHz}). These dependencies also arise from the EMRI velocity and the ejecta mass, but are more complex than those of the duration, as they are influenced by the adiabatic expansion of the ejecta. 
Furthermore, the QPE temperature $T_{\rm QPE}$ shows a weak dependence on the inclination, but a strong sensitivity to the accretion rate (see bottom panel of Figures~\ref{fig:color_2mHz} and~\ref{fig:color_0.2mHz}). The QPE temperature is determined by the QPE luminosity and the thermalization efficiency, making the origin of its parameter dependence more complex.

The primary differences between the small- and large-separation cases are a shorter duration and a higher temperature in the small separation case. If the EMRI is closer to the central SMBH, the EMRI crosses the disk at a higher velocity and the surface density at the crossing point is higher. The large velocity not only shortens the diffusion timescale, but also reduces the radius of influence. The latter tends to increase the ejecta mass, but this is outweighed by the reduction in radius of influence caused by the higher EMRI velocity. As a result, the ejecta becomes less massive, resulting in a shorter diffusion timescale.

\subsection{Spectrum of the QPE and the TDE disk} \label{subsec:Spec}
The observability of the QPE is not only determined by the QPE observables discussed above, but also by whether the QPE signals are hidden by the emission due to the disk. For the QPE spectra, we assume a free-free spectrum, which is 
\begin{align}
    L_{\nu} \propto \nu^{0}{\rm e}^{-h\nu/k_{\rm B}T_{\rm QPE}},
\end{align}
where $h$ is the Planck constant, and $\nu$ is the photon frequency. The normalization is determined such that the peak luminosity coincides with $L_{\rm QPE}$. We apply this formula even if $T_{\rm QPE}$ exceeds $50\,{\rm keV}$~\footnote{For high QPE temperatures, especially when $T_{\rm QPE}\gtrsim 100\,{\rm keV}$, the spectrum may be modified by processes such as Comptonization or pair production. However, since our interest lies not in the precise spectral shape but in whether the QPE luminosity exceeds the underlying disk luminosity at a given temperature, we adopt the simplifying assumption that the QPE spectrum follows a free-free emission spectrum even at high temperatures.}. In contrast, the disk emission is given by
\begin{align}
    \nu L_{\nu} &= \int_{R_{\mathrm{in}}}^{\infty} \nu B_{\nu}(T_{\mathrm{eff}}(R)) 2\pi R\mathrm{d}R,
\end{align}
where $B_{\nu}(T)$ is the Planck function and $T_{\rm eff}(R)$ is the effective temperature of the disk at radius $R$, given by \citep[e.g.,][]{Kato+2008}
\begin{align}
    T_{\rm eff}(R) &= \frac{3GM_{\bullet}\dot{M}}{8\pi \sigma_{\rm SB}r^3}g,
\end{align}
where $\sigma_{\rm SB}$ is the Stefan-Boltzmann constant. The inner edge of the disk, $R_{\rm in}$, is set to the innermost stable circular orbit in Schwarzschild spacetime, which is located at $R_{\rm ISCO}=6R_{\rm g}$. On the other hand, the outer edge, $R_{\rm out}$, is approximated as extending to infinity. This assumption is reasonable, since a few years after the TDE---the phase of interest in this study---the disk is expected to have expanded sufficiently, making its outer structure largely irrelevant for the high-energy portion of the spectrum. 

\begin{figure}[t]
  \begin{center}
  \includegraphics[keepaspectratio, scale=0.35]{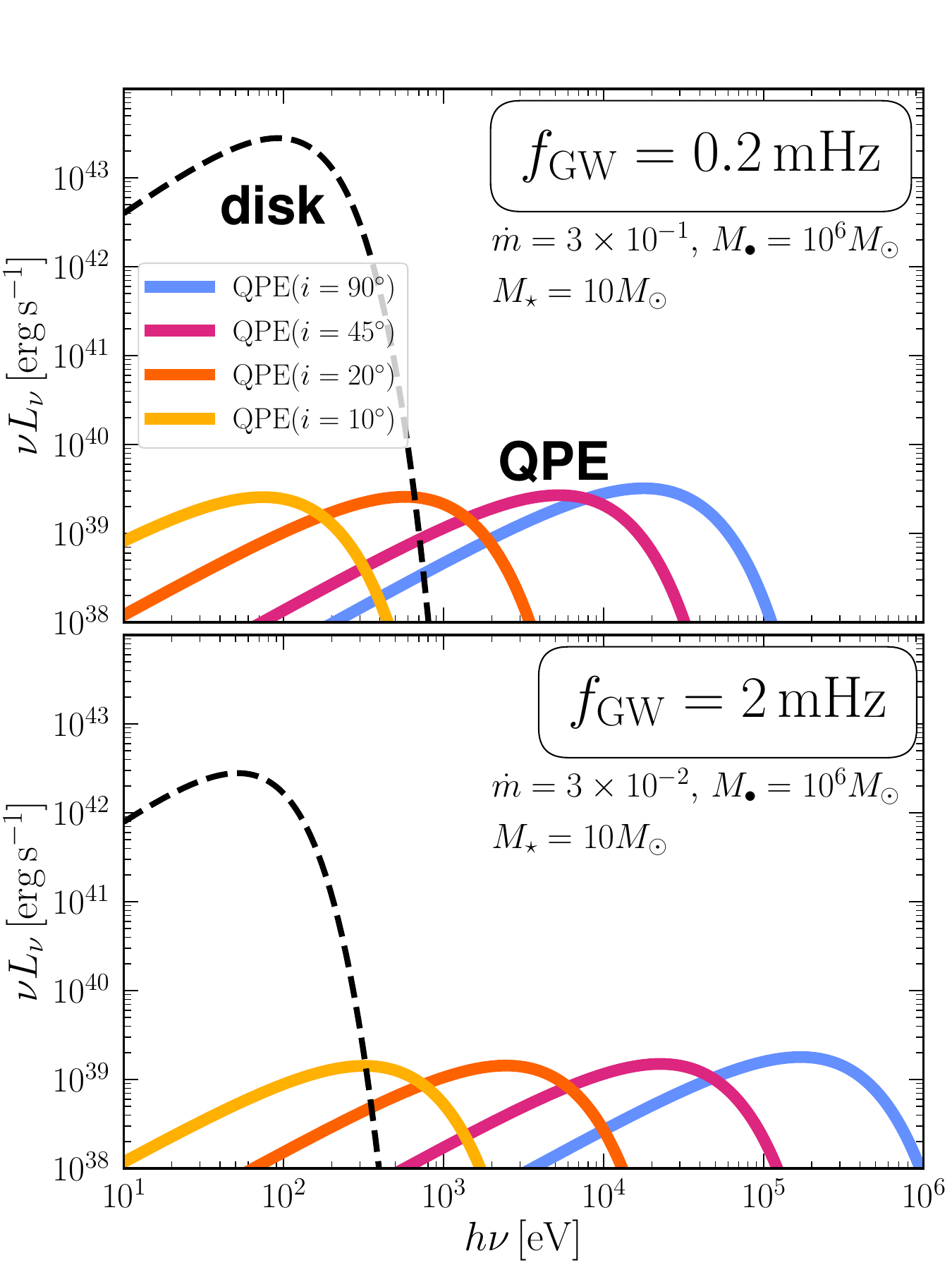}
  \end{center}
  \caption{Spectra of the disk (black dashed) and QPE for $f_{\rm GW}=0.2\,{\rm mHz}$ ($a=50R_{\rm g}$; upper panel) and $f_{\rm GW}=2\,{\rm mHz}$ ($a=10R_{\rm g}$; lower panel). Different colors represent different inclination angles between the disk and the EMRI orbital plane. We adopt $\dot{m}=3\times10^{-1}$ for $f_{\rm GW}=0.2\,{\rm mHz}$ and $\dot{m}=3\times10^{-2}$ for $f_{\rm GW}=2\,{\rm mHz}$, respectively. }
  \label{fig:Spec}
\end{figure}

Figure~\ref{fig:Spec} shows the QPE spectra for both $f_{\rm GW}=0.2\,{\rm mHz}$ and $f_{\rm GW}=2\,{\rm mHz}$. We set the accretion rate to $\dot{m}=3\times10^{-1}$ for $f_{\rm GW}=0.2\,{\rm mHz}$ and $\dot{m}=3\times10^{-2}$ for $f_{\rm GW}=2\,{\rm mHz}$. Because of the significantly higher disk luminosity, the QPE signal is entirely obscured when $k_{\rm B}T_{\rm QPE}\lesssim 1\,{\rm keV}$, which corresponds to the case $i<20^{\circ}$ with $f_{\rm GW} = 0.2\, {\rm mHz}$. If the separation is smaller, the QPE temperature becomes much higher, and thus $k_{\rm B}T_{\rm QPE}\gtrsim 10^{3}\,{\rm eV}$ in all cases. However, when the inclination is large, the temperature becomes excessively high, shifting the emission into the hard X-ray band. This tendency is more pronounced at higher accretion rates. To summarize, observation of QPEs is possible in the X-ray band, and depending on the parameters, it falls into either the soft or hard X-ray band\footnote{However, QPEs with higher temperatures, particularly those with $k_{\rm B}T_{\rm QPE}>30\,{\rm keV}$, are difficult to observe because of the reduced sensitivity of current X-ray observatories in this energy range. }.

\section{Prospects of joint EM-GW observation} \label{sec:observability}

In the previous section, we described the mHz-QPE model. Here, we estimate the event rate of multi-messenger observations of mHz-QPEs with our model. We first estimate the detection horizon of the future X-ray and GW detectors, which we take to be \textit{Lynx} and LISA. Then, we provide an order-of-magnitude estimate of the mHz-QPE event rate over the LISA lifetime. We also discuss the prospects for testing gravity theories through multi-messenger observations of QPEs.

\subsection{Detection horizon of X-ray} \label{subsec:Xray_observability}
In previous sections, we showed that QPEs are observable in the X-ray band, and depending on the parameters, the emission can fall into either the soft or the hard X-ray band. Several next-generation X-ray detectors are currently under development. For example, \textit{Athena}~\citep[$0.2-15\,{\rm keV}$;][]{Nandra+2013}, \textit{Advanced X-ray Imaging Satellite} (\textit{AXIS})~\citep[$0.5-10\,{\rm keV}$;][]{Mushotzky+2019}, and \textit{Lynx}~\citep[$0.2-10\,{\rm keV}$;][]{Gaskin+2019}. Among these, \textit{Lynx} is scheduled to begin operations in the mid-2030s, coinciding with the planned launch of LISA. Hereafter, we adopt \textit{Lynx} as the fiducial X-ray detector to estimate the detectability. 

\begin{figure}[t]
  \begin{center}
  \includegraphics[keepaspectratio, scale=0.43]{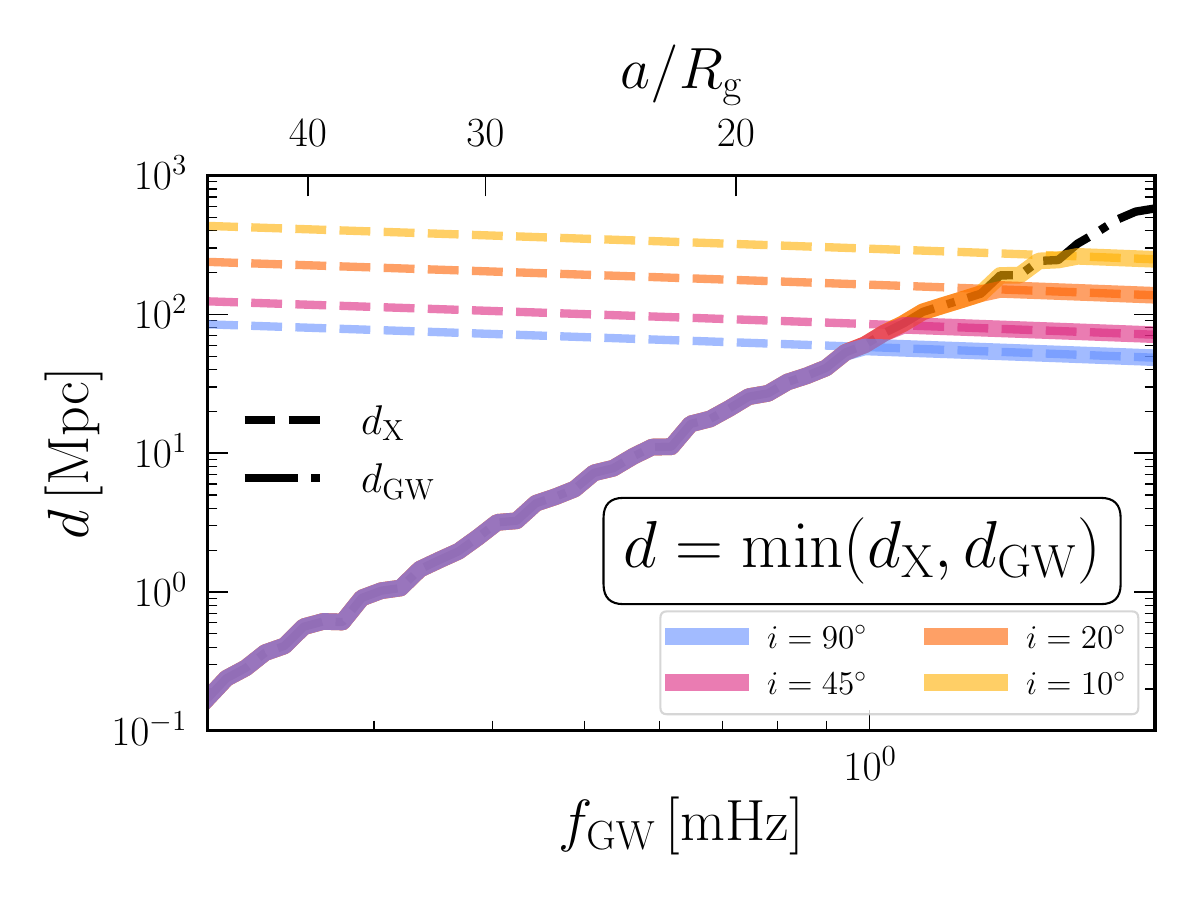}
  \end{center}
  \caption{The multi-messenger detection horizon as a function of GW frequency (or the orbital separation between the central SMBH and the EMRI). Dashed lines show the source distances detectable by a future X-ray observatory, \textit{Lynx}, $d_{\rm X}$ (see Equation~\eqref{eq:Xray_horizon}). Different colors correspond to different inclination angles between the disk and the EMRI orbital plane. Dot-dashed lines show the detection horizon of the GW produced by $10^6M_{\odot}$-$10M_{\odot}$ systems, $d_{\rm GW}$ (see Figure~\ref{fig:GW}). The multi-messenger detection horizon is determined by the minimum of the X-ray and the GW detection horizon, $d={\rm min}(d_{\rm X},d_{\rm GW})$, and this is plotted by solid lines. The accretion rate is fixed to $\dot{m}=3\times10^{-1}$, but the results are not changed significantly for $\dot{m}=3\times10^{-2}$. }
  \label{fig:mHz_distance}
\end{figure}

The sensitivity of \textit{Lynx} is given by 
\begin{align}
    \label{eq:lynx_sensitivity}
    f_{\mathrm{lim}}(t_{\rm exp}) = 1.1\times10^{-16}\,\mathrm{erg}\,\mathrm{s}^{-1}\,\mathrm{cm}^{-2}\,
    \left(\frac{t_{\mathrm{exp}}}{10^3\,\mathrm{s}}\right)^{-1/2},
\end{align}
where $t_{\mathrm{exp}}$ is the exposure time~\citep{Lops+2023}. When observing QPE, the signal lasts for $\sim t_{\rm QPE}$. Thus, we set the exposure time to $t_{\rm exp}=t_{\rm QPE}$ in our estimation. The QPE signal can be observed when the flux from the event exceeds $f_{\rm lim}$. Therefore, the maximum distance at which QPEs can be detected by \textit{Lynx} is given by
\begin{align}
    \label{eq:Xray_horizon}
    d_{\rm X} &= \left(\frac{L_{\mathrm{QPE}}}{4\pi f_{\rm lim}}\right)^{1/2}.
\end{align}
This is shown by dashed lines in Figure~\ref{fig:mHz_distance}. The X-ray signals from QPEs are easier to detect at lower frequencies due to their longer duration. In addition, systems with lower inclination angles are more easily detectable, as their longer durations improve the signal-to-noise ratio.

\subsection{Detection horizon of Gravitational Waves} \label{subsec:GW_observability}
We evaluate the detectability of GWs emitted by QPE systems, under the assumption that they originate from EMRIs. 
The detectability is quantified by the signal-to-noise ratio (SNR) of the GW signal as observed by LISA. 

The optimal SNR for the GW strain signal $h_{\rm GW}(t)$ is given by~\citep{Maggiore2007}
\begin{align}
    \rho_{\rm opt}=\sqrt{(h_{\rm GW}|h_{\rm GW})}\,.
    \label{eq:SNR_opt}
\end{align}
where the inner product in the frequency domain is defined as
\begin{align}
    (A|B) = 4 {\rm Re} \left[ \int^{f_{\rm max}}_{f_{\rm min}}\frac{\tilde{A}^*(f)\cdot\tilde{B}(f)}{S_n(f)}df \right]\,,
\end{align}
and $S_{n}(f)$ is the single-sided power spectral density of the detector noise. $f_{\rm max}$ and $f_{\rm min}$ are the maximum and the minimum frequencies spanned during the binary evolution due to GW radiation. As mentioned before, although the EMRI interacts with the disk, its orbital evolution is predominantly governed by GW radiation in the case of inclined orbits. In the following, we treat $f_{\rm max}$ as a free parameter and determine $f_{\rm min}$ using the radiation reaction formula, assuming a mission duration of $T_{\rm LISA} = 4\,{\rm yr}$ for LISA.

As a simplifying assumption, we consider a quasi-circular orbit for the inspiraling binary system.
The strain signal in the time domain is given by
\begin{align}
    h_{\rm GW}(t) = \frac{1}{2}\left[F^{+}(t)(1+\cos^2{\iota})-2iF^{\times}(t)\cos{\iota}\right]\nonumber \\
    \times\frac{4 G{\cal M}}{d_L} \left( 2\pi G {\cal M} 
{\mathcal F} \right)^{2/3}e^{-2i\Phi}\,,
\end{align}
where $F^{+, \times}(t)$ are the antenna pattern functions of LISA for the tensor modes, ${\mathcal F}$ is the orbital frequency, and $\Phi$ is the orbital phase~\citep{Hawking&Israel1987, Droz+1999}. Here, $\mathcal{M} = \eta^{3/5}m$ denotes the chirp mass with the total mass $m = M_{\bullet} + M_{\star}$ and the symmetric mass ratio $\eta = M_{\bullet}M_{\star}/m^2$, $d_{L}$ represents the luminosity distance to the source, and $\iota$ is the inclination angle of the binary orbital plane relative to the line of sight.

Applying the stationary phase approximation, the GW signal in the frequency domain takes the form 
\begin{align}
    \tilde{h}_{\rm GW}(f) = -\left[F^{+}(t(f))(1+\cos^2{\iota})-2iF^{\times}(t(f))\cos{\iota}\right]  
\nonumber\\
\times\sqrt{\frac{5\pi}{96}}\frac{(G {{\cal M}})^2}{d_L}
\left( u^{(2)} \right)^{-7/2}
e^{-i\Psi_\text{GR}^{(2)}}\,,
\label{eq:h_f}
\end{align}
where the reduced second harmonic frequency is defined by $u^{(2)}:=(\pi G\mathcal{M} f)^{1/3}$ and $\Psi_{\rm GR}^{(2)}$ is the inspiral GW phase of the second harmonic~\citep{Cutler1998, Berti+2005, Maggiore2007, Takeda+2019}. Here, $t(f)$ denotes the time to coalescence, as a function of the GW frequency~\citep{Damour+2001, Maggiore2007}.

In the low-frequency regime, the signal can be approximated as quasi-monochromatic. Hence, by averaging over the extrinsic angular parameters in Equation~\eqref{eq:h_f} and assuming $f_{\rm max}=f_{\rm GW}$, we estimate the optimal SNR as
\begin{align}\label{eq:optimalSN}
    \rho_{\rm opt} = 8 \left(\frac{\mathcal{M}}{250M_{\odot}}\right)^{5/3}\left(\frac{0.8\ {\rm Mpc}}{d_L}\right)\left(\frac{f_{\rm GW}}{0.2\ {\rm mHz}}\right)^{2/3}\nonumber \\
    \times\left(\frac{T_{\rm LISA}}{4\ {\rm yrs}}\right)^{1/2}\left(\frac{10^{-17}\ {\rm Hz^{-1/2}}}{\sqrt{S_n(f_{\rm GW})}}\right)\,.
\end{align}

To evaluate the detectability of QPE systems via GWs, we fix the mass of the central BH to $M_{\bullet} = 10^6 M_\odot$ (see Section~\ref{subsec:parameter}), and vary the mass of the compact secondary object $M_{\star}$ in the range $1$-$30 M_\odot$. We consider quasi-circular orbits with no spin. The upper cutoff GW frequency is scanned over the range $0.2$-$2\,\mathrm{mHz}$. 
For each mass and frequency configuration, we assume a four-year observation by LISA, assuming the design sensitivity curve and a heliocentric orbit~\citep{Amaro-Seoane+2017}. 

\begin{figure}[t]
  \begin{center}
  \includegraphics[keepaspectratio, scale=0.45]{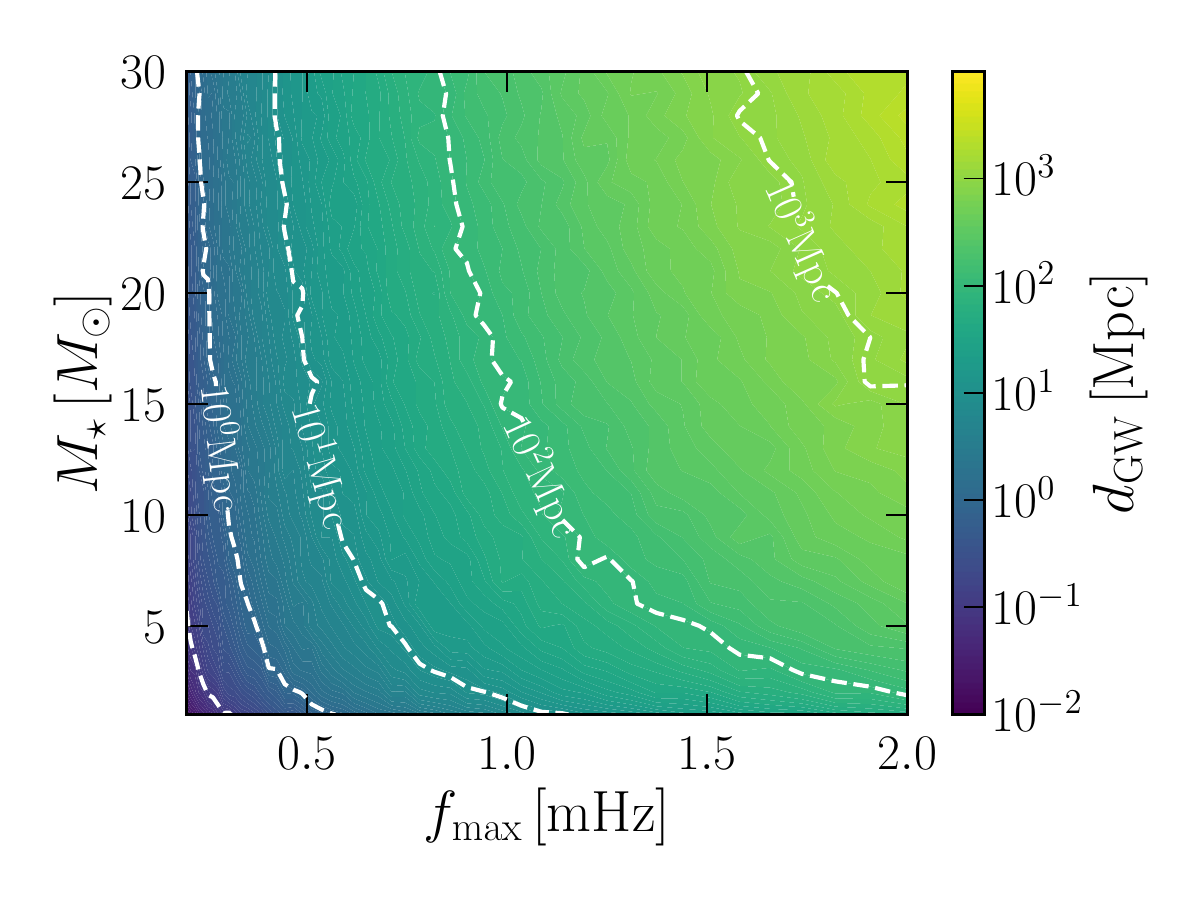}
  \end{center}
  \caption{Detection horizon $d_{\rm GW}$ for typical QPE-EMRI systems detectable with SNR $\rho_{\rm opt}=8$ by LISA as a function of secondary mass and upper cutoff GW frequency. The primary mass is fixed to $10^6M_{\odot}$. A four-year mission duration is assumed for LISA, along with its design sensitivity and a heliocentric orbit.}
  \label{fig:GW}
\end{figure}

To account for variation due to angular parameters (inclination, sky location, and polarization angle), we simulate 100 randomly sampled configurations from a uniform distribution on the sphere, and compute the corresponding SNRs using Equations.~\eqref{eq:SNR_opt} and \eqref{eq:h_f}. The median of these SNRs is then used as a representative value. Figure~\ref{fig:GW} shows the typical source distance $d_{\rm GW}$ at which a QPE system is detectable with an SNR of 8 by LISA, plotted as a function of the secondary mass and upper cutoff frequency.

The detection horizon for GWs is also shown by dot-dashed lines in Figure~\ref{fig:mHz_distance}. At lower frequencies, the X-ray signals from QPEs are easier to detect~(see Section~\ref{subsec:Xray_observability}), while the GW signals are relatively weak. Consequently, the detection horizon in this regime is primarily limited by the sensitivity of GW observatories. As the frequency increases, GW signals become stronger, whereas the X-ray signals become shorter in duration, making them more challenging to observe. In this higher-frequency regime, the detection horizon is instead determined by the sensitivity of X-ray observatories.

\subsection{Simultaneous Observability} \label{subsec:simltaneous}
With the above detection horizons, we can estimate the number of mHz-QPE events within the LISA lifetime $T_{\rm LISA}$, which we denote $\cal{N}_{\rm mHz-QPE}$ by 
\begin{align}
    {\cal N}_{\rm mHz-QPE} \sim {\cal N}\dot{N}_{\rm TDE}T_{\rm LISA}.
\end{align}
Here, ${\cal N}$ is the total number of galaxies hosting EMRI within the detection horizon $d={\rm min}(d_{\rm X},d_{\rm GW})$~(see Figure~\ref{fig:mHz_distance}), and $\dot{N}_{\rm TDE}$ is the event rate of TDE per galaxy. The TDE rate $\dot{N}_{\rm TDE}$ has been estimated in the various studies~\citep[e.g.,][]{Syer&Ulmer1999,Wang&Merritt2004,Stone&Metzger2016}. In this paper, we adopt the result from~\citet{Stone&Metzger2016}, which gives the TDE rate as: 
\begin{align}\label{eq:TDErate}
    \dot{N}_{\mathrm{TDE}} &= 1.9\times10^{-4}\,\mathrm{yr}^{-1}\,\left(\frac{M_{\bullet}}{10^6M_{\odot}}\right)^{-0.404}.
\end{align}

To estimate ${\cal N}$, it is required to know the EMRI formation rate per galaxy, denoted as $\Gamma_{\mathrm{EMRI}}$. Several previous studies have estimated the EMRI formation rate~\citep[e.g.,][]{Amaro-Seoane&Preto2011,Babak+2017,Pan&Yang2021}, with results suggesting $\Gamma_{\mathrm{EMRI}}\sim10^{-5}-10^{-7}\,\mathrm{yr}^{-1}$~\citep[see also][]{Amaro-Seoane2018}. In this study, we adopt a more optimistic value of $\Gamma_{\mathrm{EMRI}}=10^{-5}\,\mathrm{yr}^{-1}$, assuming that the presence of an accretion disk enhances the efficiency of EMRI formation. This assumption is reasonable because some QPE host galaxies exhibit narrow-line regions, indicating past activity linked to the presence of an accretion disk. Now, the total EMRI rate for a source at distance $d$ is given by 
\begin{align}
    \Gamma_{\mathrm{tot}} &= \Gamma_{\mathrm{EMRI}}\times n_{\bullet}\times (4\pi/3)d^3, 
\end{align}
where $n_{\bullet}$ is the number density of SMBHs, expressed as 
\begin{align}\label{eq:SMBHrate}
    \frac{\mathrm{d}n_{\bullet}}{\mathrm{d}\log{M_{\bullet}}} &= 2\times10^{-3}
    \left(\frac{M_{\bullet}}{3\times10^6M_{\odot}}\right)^{0.3}\,\mathrm{Mpc}^{-3},
\end{align}
~\citep[e.g.,][]{Babak+2017}. For a fiducial SMBH mass of $M_{\bullet}=10^6M_{\odot}$, the number density is $n_{\bullet}\sim10^7\,\mathrm{Gpc}^{-3}$. The source distance is determined by $d={\rm min}(d_{\rm X},d_{\rm GW})$. To evaluate the number of detectable sources, we also need to consider the time each separation lasts. In our framework, the orbital evolution of the EMRIs is governed by GWs. Therefore, the staying time can be approximated by the coalescence time due to GW emission, which is $t_{\mathrm{GW}}\simeq(5/256)(c^{5}a^{4}/(G^{3}M_{\bullet}^2M_{\star}))$. Finally, the detectable number of sources is 
\begin{align}
    \mathcal{N} &= \Gamma_{\mathrm{tot}}t_{\mathrm{GW}},
\end{align}
which represents the number of galaxies that host EMRIs. 

\begin{figure}[t]
  \begin{center}
  \includegraphics[keepaspectratio, scale=0.43]{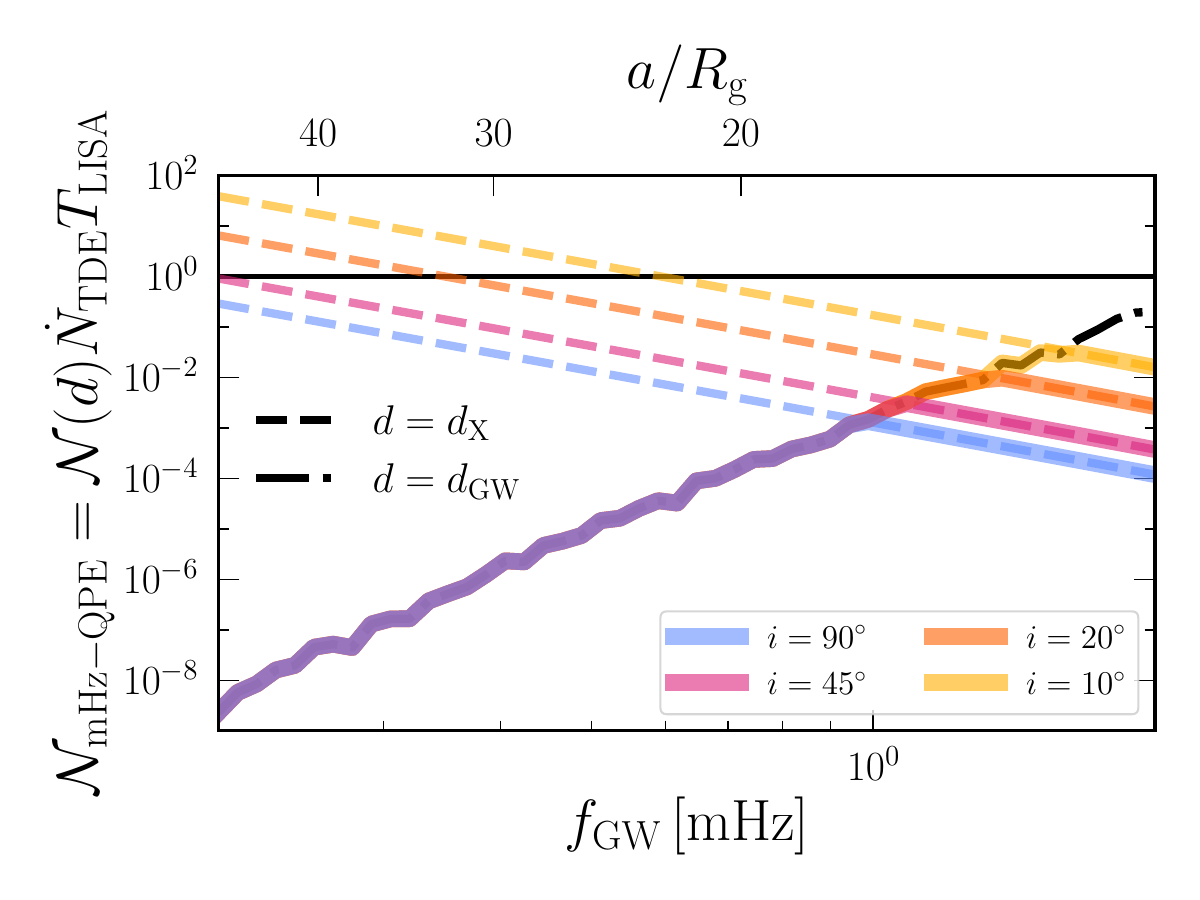}
  \end{center}
  \caption{The number of mHz-QPE detectable in a 4-year mission by LISA. Different colors correspond to different inclination angles. The accretion rate is fixed to $\dot{m}=3\times10^{-1}$, and the results are not changed significantly for $\dot{m}=3\times10^{-2}$. }
  \label{fig:mHz_number}
\end{figure}

Figure~\ref{fig:mHz_number} shows the number of mHz-QPEs for different inclination angles. We observe that the number peaks around $1\,{\rm mHz}$, depending on the inclination. At frequencies above this peak, the number of observable events is limited by X-ray observability, as shorter QPE durations make detection more challenging.  Conversely, in the lower frequency range, it is easier to observe QPEs in the X-ray due to their longer durations, with the expected number of detectable events ranging from 1 to 100. However, in this frequency range, the simultaneous detectability is ultimately constrained by GW sensitivity. EMRIs in the sub-mHz band are challenging to detect with LISA because of their lower signal. 

As discussed in Section~\ref{subsec:Xray_observability}, QPEs with lower inclinations have an advantage in observation. This fact is reflected in Figure~\ref{fig:mHz_number}, where the peak number increases as the inclination is lowered. 

Up to this point, we have fixed the EMRI mass $M_{\star}$ and the SMBH mass $M_{\bullet}$, but the dependence on these parameters can be straightforwardly derived. For a fixed GW frequency $f_{\rm GW}$, the X-ray detection horizon scales as $d_{\rm X} \propto M_{\star}^{7/12}M_{\bullet}^{-1/18}$, which follows from substituting Equations.~\eqref{eq:f_gw}, \eqref{eq:duration}, \eqref{eq:luminosity}, and \eqref{eq:lynx_sensitivity} into Equation~\eqref{eq:Xray_horizon}. Meanwhile, the GW detection horizon scales as $d_{\rm GW} \propto M_{\star}M_{\bullet}^{2/3}$, as given by Equation~\eqref{eq:optimalSN}. Noting that $t_{\rm GW} \propto M_{\star}^{-1}M_{\bullet}^{-2/3}$ and using Equations.~\eqref{eq:TDErate} and~\eqref{eq:SMBHrate}, the expected detection number of mHz-QPEs is estimated as 
\begin{align}
    \mathcal{N}_{\rm mHz-QPE} \propto
    \begin{cases}
    \Gamma_{\rm EMRI} M_*^{3/4} M_\bullet^{-0.94}~, & (\text{X-ray})~,\\
    \Gamma_{\rm EMRI} M_*^{2} M_\bullet^{0.63}~, &  (\text{GW})~.
    \end{cases}
\end{align}
Here, we retain $\Gamma_{\rm EMRI}$ to account for its uncertain dependence on $M_\bullet$ and $M_*$. This scaling indicates that increasing the EMRI mass boosts the maximum event rate. For example, if one considers a $30M_{\odot}$ companion, the rate is enhanced by one order of magnitude, if the EMRI rate is the same as the $10M_{\odot}$ companion case. On the other hand, variations in the SMBH mass have a relatively weaker impact. 

To summarize, the maximum number of simultaneously detectable sources is $\mathcal{N}_{\rm mHz-QPE} \lesssim 1$, which occurs at the lowest possible inclination. Therefore, we conclude that mHz-QPEs detectable by both GW and X-ray are rare in our framework. 

\subsection{Possibility to constrain the propagation speed of gravitational waves in the mHz band}

In the last section, we found that the simultaneous GW and EM observations of QPEs might be rare. However, if we detect such rare events, they may offer a new channel for probing fundamental physics, particularly by enabling direct constraints on the propagation speed of GWs in the mHz band. We roughly estimate the constraint on the GW propagation speed if we could detect a QPE with both GW and EM.

LISA has been identified as a powerful instrument for testing fundamental physics, including the propagation properties of GWs over cosmological distances~\citep{Arun+2022}.
A direct constraint on the GW propagation speed was first achieved in the high-frequency band through the observation of GW170817 and its associated gamma-ray burst, which yielded a stringent bound of $\Delta v/c \lesssim 10^{-15}$ in the $\sim$100 Hz band~\citep{Abbott+2019}.
In contrast, only a few proposals exist in the case of LISA. One example is model-dependent tests that constrain possible frequency-dependent deviations through waveform distortions~\citep{Baker+2022}. The other example is the multimessenger detection of a white dwarf binary in the galaxy. We would like to mention that QPE can also provide a tight constraint on the GW propagation speed.

If QPEs are indeed generated by EMRIs and produce both GW and electromagnetic (EM) signals observable in coincidence, they would provide a rare opportunity to directly constrain the GW propagation speed in the mHz band.  
In such a case, the arrival time delay $\Delta t$ between the GW and EM signals could be measured by comparing the phase of the GW waveform with the peak timing of the QPE in the EM light curve.  
The relative speed difference is then estimated by ${\Delta v}/{c} = (c/d_{L}) \Delta t$ where $d_{L}$ is the luminosity distance to the source.  
We assume the EM timing precision to be limited by the duration of the QPE signal, as given by Equation~\eqref{eq:duration}, while the GW timing uncertainty can be estimated from the phase uncertainty as $ \Delta t_{\rm GW} \sim \Delta \phi / (2\pi f)$.  
Assuming a phase accuracy of $ \Delta \phi \sim 0.1 $ radians and a typical GW frequency $ f \sim 2\,{\rm mHz} $, we obtain $ \Delta t_{\rm GW} \sim 8\,{\rm s} $, which is longer than the typical QPE duration time (see Figure~\ref{fig:color_2mHz}).  
Assuming that the arrival time delay is dominated by the GW phase uncertainty, one can constrain the relative speed difference to the level of $\Delta v / c \sim 10^{-15}$ for a typical source distance of $ d_{L} = 100\,{\rm Mpc} $.
We note that in the case of white dwarf binaries, the speed of light can be constrained using eclipse timing, which fixes the viewing geometry~\citep{Larson&Hiscock2000, Cutler+2003, Littenberg&Cornish2019}.  
In contrast, QPEs do not necessarily provide such geometric constraints with our simplified model.  
Therefore, the degeneracy between propagation delay and intrinsic phase offset between the GW and EM signals must be taken into account, and would likely weaken the achievable bound in practice. We leave a detailed study in this direction as a future work.

The constraint on the GW speed from QPE would not only be independent of constraints obtained by ground-based detectors, but would also probe a qualitatively distinct frequency regime.  
In particular, various modified gravity theories predict frequency-dependent GW propagation speeds, including models involving Lorentz invariance violation, massive gravity, and certain effective field theory extensions of general relativity~\citep{deRham&Melville2018}.

\section{Other possibilities} \label{sec:other}
In the discussion so far, we have assumed that the EMRI is a stellar-mass BH and that the accretion disk originates from a TDE. However, alternative scenarios are also possible. The EMRI could instead be a main-sequence star or a more compact object, such as a white dwarf (WD) or neutron star (NS). We explore the joint GW and EM detectability of such systems in Section~\ref{subsec:star_QPE} and~\ref{subsec:WDNS}. In addition, the accretion disk itself might be associated with an active galactic nucleus (AGN), rather than a TDE remnant, which we discuss in Section~\ref{subsec:AGNQPE}.

\subsection{Simultaneous detectability of star-disk QPEs by GW and EM} \label{subsec:star_QPE}
In this section, we discuss whether star-disk QPEs can be simultaneously detected through GW and X-ray observations. When a star approaches a SMBH, it experiences tidal forces. If the tidal forces exceed the star's self-gravity, the star is disrupted. This disruption occurs within the tidal radius, given by 
\begin{align}
    R_{\mathrm{T}} &= \left(\frac{M_{\bullet}}{M_{\star,\mathrm{TDE}}}\right)^{1/3}R_{\star,\mathrm{TDE}} \nonumber \\
    &\simeq 47R_{\mathrm{g}}\,
    \frac{M_{\bullet,6}^{1/3}\mathcal{R}_{\star,\mathrm{TDE}}}{m_{\star,\mathrm{TDE}}^{1/3}},
\end{align}
where $M_{\star,\mathrm{TDE}}=M_{\odot}m_{\star,\mathrm{TDE}}$ and $R_{\star,\mathrm{TDE}}=\mathcal{R}_{\star,\mathrm{TDE}}$ are the mass and radius of the disrupted star, respectively~\citep{Rees1988,Phinney1989}. The corresponding GW frequency at the tidal radius is 
\begin{align}
    f_{\mathrm{GW,TDE}} &= \frac{1}{\pi}\left(\frac{GM_{\bullet}}{R_{\mathrm{T}}^{3}}\right)^{1/2} \nonumber \\
    &\simeq 0.20\,\mathrm{mHz}\,
    \frac{m_{\star,\mathrm{TDE}}^{1/2}}{\mathcal{R}_{\star,\mathrm{TDE}}^{3/2}}.
\end{align}
Therefore, the star is tidally disrupted near the lower end of the LISA band. This implies that GWs from star-disk QPE systems are difficult to detect with LISA, as the star does not survive long enough to emit within the sensitivity frequency range.

\subsection{Other compact object cases; white dwarfs and neutron stars} \label{subsec:WDNS}
In this section, we discuss whether compact objects other than BHs, such as WDs and NSs, can produce observable QPEs. For BHs, the radius of influence is defined by the HL radius, as given in Equation~\eqref{eq:HL_radius}. In contrast, for WDs and NSs with stellar radii $R_{\star}$, the relevant influence radius is $R_{\rm inf}={\rm max}(R_{\star},R_{{\rm HL},\star})$. Typically, the radius of WDs is $\sim 0.01R_{\odot}$, which is larger than their HL radius. On the other hand, NSs have radii of $\sim 10\,{\rm km}$, which are smaller than their HL radii. Therefore, the influence radius is $R_{\star}$ for WDs and $R_{{\rm HL},\star}$ for NSs. 

The WDs have influence radii determined by their physical size, $\sim0.01R_{\odot}$. The QPE observables in the WD case can be evaluated by $R_{\inf}=R_{\star}$. According to Equations.~\eqref{eq:duration} and \eqref{eq:luminosity}, the QPE duration and luminosity scale with $\propto R_{\star}$ and $\propto R_{\star}^{1/3}$, respectively. Since the WD radius is generally larger than the HL radius of a $10M_{\odot}$ BH, WD-induced QPEs are expected to be more favorable for X-ray detection. However, as shown in Figure~\ref{fig:GW}, WDs are significantly less massive than BHs, and thus GW signals from WD-EMRI systems are detectable only at relatively small distances. As a result, the event rate for GW detections from such systems is expected to be very low. Therefore, although X-ray detection of WD-QPEs may be feasible, simultaneous detection via both GW and X-ray is likely to be rare. 

In contrast, the influence radius for NSs is the HL radius ---just as in the BH case---the only significant difference between NSs and BHs lies in their masses. In our framework, the BH mass is assumed to be $\sim 10M_{\odot}$, while the NSs typically have masses around $\sim1M_{\odot}$. As shown in Section~\ref{sec:emission}, QPE signals produced by lower-mass objects are not only fainter but also shorter in duration, making them more challenging to detect.

\subsection{Detectability of AGN-QPE} \label{subsec:AGNQPE}
In this section, we discuss the detectability of mHz-QPEs associated with AGN, which has been proposed in~\cite{Lyu+2025}. As mentioned in Section~\ref{subsec:simltaneous}, the event rate of TDEs is relatively low, making the formation of disks in ``EMRI+disk'' systems via TDEs rare. In contrast, AGN disks are more ubiquitous, as AGNs are present in approximately $\gtrsim 1\%$ of galaxies. Therefore, QPEs associated with AGNs are more favorable in terms of occurrence rate. 

The evaluation of QPE observables follows the same methodology as in Section~\ref{sec:emission}, the case where the disk is formed by a TDE. Consequently, the QPE spectra are expected to be similar to those shown in Figure~\ref{fig:Spec}. However, the underlying disk emission differs significantly from that of TDE disks. TDE disks typically have a thermal spectrum in the UV to soft X-ray band and lack significant higher-energy emission components, as implied from observation~\citep[e.g.,][]{Gezari2021}. In contrast, the spectrum of AGN disks has higher-energy emission components originating from a hot corona as well as the disk blackbody emission~\citep[e.g.,][]{Liu+2002,Liu+2003,Cao2009,Qiao&Liu2017,Qiao&Liu2018}. This higher-energy emission can obscure QPE signals, especially when the QPE spectrum lies in the $0.1-10\,{\rm keV}$ band. As a result, even if the QPEs themselves are intrinsically bright and in the favorable X-ray band, their detection becomes challenging due to contamination from AGN emission.

\section{Conclusion} \label{sec:conclusion}
We have investigated the potential for simultaneous detection of quasi-periodic eruptions (QPEs) in both gravitational wave (GW) and electromagnetic (EM) signals, based on the EMRI+disk model in which a stellar-mass object periodically interacts with an accretion disk around a supermassive black hole (SMBH). We assume that the companion orbiting the SMBH is a stellar-mass black hole (BH) since such systems can emit both GW and EM signals. We also assume that the disk is formed through a tidal disruption event (TDE), where a star is torn apart by the tidal force of the SMBH. 

Our analysis reveals that the observables of QPEs, such as their duration, luminosity, and temperature, are strongly dependent on the system's inclination angle and accretion rate. In particular, low-inclination systems with moderate accretion rates tend to produce longer, less dim X-ray flares that are more favorable for observation. 

The QPE emissions from the system with an orbital period around $0.1\,{\rm mHz}$ and low inclination angles may be obscured by the underlying disk emission due to their lower temperatures. This makes medium-inclination systems more favorable for X-ray detection in this frequency range. Under the assumption of future X-ray observatories like \textit{Lynx}, the expected number of detections in this band remains $\gtrsim{\cal O}(1)$, indicating that mHz-QPEs could indeed be observed in the future. However, since GW signals from these low-frequency systems are too weak and fall below the sensitivity of future space-based GW observatories like LISA, they are unlikely to serve as promising targets for simultaneous GW and EM detections. 

In contrast, systems with an orbital period around $1\,{\rm mHz}$ are more promising targets for joint detections, as their spectra are less likely to be obscured by the disk emission, even for the low inclination systems, and their GW signals are strong enough to detect distant sources. However, our calculations suggest that the number of EMRI+disk systems simultaneously detectable by both X-ray telescopes and LISA is limited. Under optimistic assumptions, the expected number of such detections over the 4-year mission is $\lesssim 1$, due to the mismatch in sensitivity windows and the rarity of suitable systems. 

Looking ahead, the situation could drastically improve with future GW detectors operating in the $\mu$Hz band~\citep{Sesana+2021, Fedderke+2022, Foster+2025}. These detectors would access lower-frequency signals corresponding to larger orbital separations and longer QPE durations, thus increasing the detectable volume. Assuming a sensitivity of $\mu$Ares~\citep{Sesana+2021} with $S_n(f=0.2\ {\rm mHz}) \sim 10^{-19}~\mathrm{Hz}^{-1/2}$, such detectors could detect typical QPE+EMRI systems out to $d_L \sim 80\ {\rm Mpc}$ for $1M_{\odot}-10^6M_{\odot}$ binaries and $d_L \sim 800\ {\rm Mpc}$ for $10M_{\odot}-10^6M_{\odot}$ binaries (see Equation~\eqref{eq:optimalSN}), although confusion noise from galactic foregrounds must be taken into account in realistic projections. This expands the accessible volume by orders of magnitude and could yield up to $\sim 10^{0-1}$ simultaneously detectable events within several years. We leave the investigation of the science opportunity with $\mu$Hz detectors as a future work.

Finally, we note some caveats regarding the EMRI+disk model. Recently, there have been a few studies that suggest this model is too simplified to predict the emission properties of QPEs accurately. For example, the disk properties are idealized. In fact, the disk structures might be more complicated. In addition, since the disk is perturbed by a single disk-crossing, it is not certain that the steady state is achieved at the next crossing. Recent work~\citep{Guo&Shen2025} suggests that the EMRI+disk model might not reproduce the existing QPE observables. Moreover, a recent hydrodynamical simulation suggests that the ejecta mass launched from the disk is lower than that estimated by the EMRI+disk model, leading to shorter duration~\citep{Tsz-Lok_Lam+2025}. 

In contrast, the longer duration and larger luminosity might be possible if some mechanisms cause the radius of influence of the BH to exceed the HL radius. For example, magnetohydrodynamical simulations suggest that such a phenomenon may occur when the accretion disk possesses a strong magnetic field oriented perpendicular to the BH's motion~\citep{Lee+2014,Nomura+2018}. It is necessary to construct a more accurate emission model for QPEs, and we plan to address this problem in future work.

\section*{Acknowledgements}
We greatly thank Takashi Hosokawa and Tatsuya Matsumoto for continuous discussions about QPEs and checking the manuscript. We also thank Kimitake Hayasaki, Shigeo S. Kimura, Ken Ohsuga, Riki Matsui, Rei Nishiura, Takafumi Kakehi, Kazumi Kashiyama, Kunihito Ioka, and Takahiro Tanaka for fruitful discussions and comments. T.S. was supported by JST SPRING, grant No. JPMJSP 110. This research was supported by a grant from the Hayakawa Satio Fund awarded by the Astronomical Society of Japan. H. O.~ as supported by JSPS KAKENHI Grant Numbers JP23H00110 and Yamada Science Foundation. 
H.T. was supported by the Hakubi project at Kyoto University and by JSPS KAKENHI Grant No. JP22K1 037.

\bibliographystyle{apj}
\bibliography{suzuguchi_reference}

\begin{thebibliography}{}
\expandafter\ifx\csname natexlab\endcsname\relax\def\natexlab#1{#1}\fi
\providecommand{\url}[1]{\href{#1}{#1}}
\providecommand{\dodoi}[1]{doi:~\href{http://doi.org/#1}{\nolinkurl{#1}}}
\providecommand{\doeprint}[1]{\href{http://ascl.net/#1}{\nolinkurl{http://ascl.net/#1}}}
\providecommand{\doarXiv}[1]{\href{https://arxiv.org/abs/#1}{\nolinkurl{https://arxiv.org/abs/#1}}}

\bibitem[{{Abbott} {et~al.}(2019){Abbott}, {Abbott}, {Abbott}, {Acernese}, {Ackley}, {Adams}, {Adams}, {Addesso}, {Adhikari}, {Adya}, {Affeldt}, {Agarwal}, {Agathos}, {Agatsuma}, {Aggarwal}, {Aguiar}, {Aiello}, {Ain}, {Ajith}, {Allen}, {Allen}, {Allocca}, {Aloy}, {Altin}, {Amato}, {Ananyeva}, {Anderson}, {Anderson}, {Angelova}, {Antier}, {Appert}, {Arai}, {Araya}, {Areeda}, {Ar{\`e}ne}, {Arnaud}, {Arun}, {Ascenzi}, {Ashton}, {Ast}, {Aston}, {Astone}, {Atallah}, {Aubin}, {Aufmuth}, {Aulbert}, {AultONeal}, {Austin}, {Avila-Alvarez}, {Babak}, {Bacon}, {Badaracco}, {Bader}, {Bae}, {Baker}, {Baldaccini}, {Ballardin}, {Ballmer}, {Banagiri}, {Barayoga}, {Barclay}, {Barish}, {Barker}, {Barkett}, {Barnum}, {Barone}, {Barr}, {Barsotti}, {Barsuglia}, {Barta}, {Bartlett}, {Bartos}, {Bassiri}, {Basti}, {Batch}, {Bawaj}, {Bayley}, {Bazzan}, {B{\'e}csy}, {Beer}, {Bejger}, {Belahcene}, {Bell}, {Beniwal}, {Bensch}, {Berger}, {Bergmann}, {Bernuzzi}, {Bero}, {Berry}, {Bersanetti}, {Bertolini}, {Betzwieser}, {Bhandare},
  {Bilenko}, {Bilgili}, {Billingsley}, {Billman}, {Birch}, {Birney}, {Birnholtz}, {Biscans}, {Biscoveanu}, {Bisht}, {Bitossi}, {Bizouard}, {Blackburn}, {Blackman}, {Blair}, {Blair}, {Blair}, {Bloemen}, {Bock}, {Bode}, {Boer}, {Boetzel}, {Bogaert}, {Bohe}, {Bondu}, {Bonilla}, {Bonnand}, {Booker}, {Boom}, {Booth}, {Bork}, {Boschi}, {Bose}, {Bossie}, {Bossilkov}, {Bosveld}, {Bouffanais}, {Bozzi}, {Bradaschia}, {Brady}, {Bramley}, {Branchesi}, {Brau}, {Briant}, {Brighenti}, {Brillet}, {Brinkmann}, {Brisson}, {Brockill}, {Brooks}, {Brown}, {Brunett}, {Buchanan}, {Buikema}, {Bulik}, {Bulten}, {Buonanno}, {Buskulic}, {Buy}, {Byer}, {Cabero}, {Cadonati}, {Cagnoli}, {Cahillane}, {Bustillo}, {Callister}, {Calloni}, {Camp}, {Canepa}, {Canizares}, {Cannon}, {Cao}, {Cao}, {Capano}, {Capocasa}, {Carbognani}, {Caride}, {Carney}, {Carullo}, {Diaz}, {Casentini}, {Caudill}, {Cavagli{\`a}}, {Cavalier}, {Cavalieri}, {Cella}, {Cepeda}, {Cerd{\'a}-Dur{\'a}n}, {Cerretani}, {Cesarini}, {Chaibi}, {Chamberlin}, {Chan}, {Chao},
  {Charlton}, {Chase}, {Chassande-Mottin}, {Chatterjee}, {Chatziioannou}, {Cheeseboro}, {Chen}, {Chen}, {Chen}, {Cheng}, {Chia}, \& {Chincarini}}]{Abbott+2019}
{Abbott}, B.~P., {Abbott}, R., {Abbott}, T.~D., {et~al.} 2019, \prl, 123, 011102, \dodoi{10.1103/PhysRevLett.123.011102}

\bibitem[{{Abramowicz} {et~al.}(1988){Abramowicz}, {Czerny}, {Lasota}, \& {Szuszkiewicz}}]{Abramowicz+1988}
{Abramowicz}, M.~A., {Czerny}, B., {Lasota}, J.~P., \& {Szuszkiewicz}, E. 1988, \apj, 332, 646, \dodoi{10.1086/166683}

\bibitem[{{Amaro-Seoane}(2018)}]{Amaro-Seoane2018}
{Amaro-Seoane}, P. 2018, Living Reviews in Relativity, 21, 4, \dodoi{10.1007/s41114-018-0013-8}

\bibitem[{{Amaro-Seoane} \& {Preto}(2011)}]{Amaro-Seoane&Preto2011}
{Amaro-Seoane}, P., \& {Preto}, M. 2011, Classical and Quantum Gravity, 28, 094017, \dodoi{10.1088/0264-9381/28/9/094017}

\bibitem[{{Amaro-Seoane} {et~al.}(2017){Amaro-Seoane}, {Audley}, {Babak}, {Baker}, {Barausse}, {Bender}, {Berti}, {Binetruy}, {Born}, {Bortoluzzi}, {Camp}, {Caprini}, {Cardoso}, {Colpi}, {Conklin}, {Cornish}, {Cutler}, {Danzmann}, {Dolesi}, {Ferraioli}, {Ferroni}, {Fitzsimons}, {Gair}, {Gesa Bote}, {Giardini}, {Gibert}, {Grimani}, {Halloin}, {Heinzel}, {Hertog}, {Hewitson}, {Holley-Bockelmann}, {Hollington}, {Hueller}, {Inchauspe}, {Jetzer}, {Karnesis}, {Killow}, {Klein}, {Klipstein}, {Korsakova}, {Larson}, {Livas}, {Lloro}, {Man}, {Mance}, {Martino}, {Mateos}, {McKenzie}, {McWilliams}, {Miller}, {Mueller}, {Nardini}, {Nelemans}, {Nofrarias}, {Petiteau}, {Pivato}, {Plagnol}, {Porter}, {Reiche}, {Robertson}, {Robertson}, {Rossi}, {Russano}, {Schutz}, {Sesana}, {Shoemaker}, {Slutsky}, {Sopuerta}, {Sumner}, {Tamanini}, {Thorpe}, {Troebs}, {Vallisneri}, {Vecchio}, {Vetrugno}, {Vitale}, {Volonteri}, {Wanner}, {Ward}, {Wass}, {Weber}, {Ziemer}, \& {Zweifel}}]{Amaro-Seoane+2017}
{Amaro-Seoane}, P., {Audley}, H., {Babak}, S., {et~al.} 2017, arXiv e-prints, arXiv:1702.00786, \dodoi{10.48550/arXiv.1702.00786}

\bibitem[{{Amaro-Seoane} {et~al.}(2023){Amaro-Seoane}, {Andrews}, {Arca Sedda}, {Askar}, {Baghi}, {Balasov}, {Bartos}, {Bavera}, {Bellovary}, {Berry}, {Berti}, {Bianchi}, {Blecha}, {Blondin}, {Bogdanovi{\'c}}, {Boissier}, {Bonetti}, {Bonoli}, {Bortolas}, {Breivik}, {Capelo}, {Caramete}, {Cattorini}, {Charisi}, {Chaty}, {Chen}, {Chru{\'s}li{\'n}ska}, {Chua}, {Church}, {Colpi}, {D'Orazio}, {Danielski}, {Davies}, {Dayal}, {De Rosa}, {Derdzinski}, {Destounis}, {Dotti}, {Du{\c{t}}an}, {Dvorkin}, {Fabj}, {Foglizzo}, {Ford}, {Fouvry}, {Franchini}, {Fragos}, {Fryer}, {Gaspari}, {Gerosa}, {Graziani}, {Groot}, {Habouzit}, {Haggard}, {Haiman}, {Han}, {Istrate}, {Johansson}, {Khan}, {Kimpson}, {Kokkotas}, {Kong}, {Korol}, {Kremer}, {Kupfer}, {Lamberts}, {Larson}, {Lau}, {Liu}, {Lloyd-Ronning}, {Lodato}, {Lupi}, {Ma}, {Maccarone}, {Mandel}, {Mangiagli}, {Mapelli}, {Mathis}, {Mayer}, {McGee}, {McKernan}, {Miller}, {Mota}, {Mumpower}, {Nasim}, {Nelemans}, {Noble}, {Pacucci}, {Panessa}, {Paschalidis}, {Pfister}, {Porquet},
  {Quenby}, {Ricarte}, {R{\"o}pke}, {Regan}, {Rosswog}, {Ruiter}, {Ruiz}, {Runnoe}, {Schneider}, {Schnittman}, {Secunda}, {Sesana}, {Seto}, {Shao}, {Shapiro}, {Sopuerta}, {Stone}, {Suvorov}, {Tamanini}, {Tamfal}, {Tauris}, {Temmink}, {Tomsick}, {Toonen}, {Torres-Orjuela}, {Toscani}, {Tsokaros}, {Unal}, {V{\'a}zquez-Aceves}, {Valiante}, {van Putten}, {van Roestel}, {Vignali}, {Volonteri}, {Wu}, {Younsi}, {Yu}, {Zane}, {Zwick}, {Antonini}, {Baibhav}, {Barausse}, {Bonilla Rivera}, {Branchesi}, {Branduardi-Raymont}, {Burdge}, {Chakraborty}, {Cuadra}, {Dage}, {Davis}, {de Mink}, {Decarli}, {Doneva}, {Escoffier}, {Gandhi}, {Haardt}, {Lousto}, {Nissanke}, {Nordhaus}, {O'Shaughnessy}, {Portegies Zwart}, {Pound}, {Schussler}, {Sergijenko}, {Spallicci}, {Vernieri}, \& {Vigna-G{\'o}mez}}]{Amaro-Seoane+2023}
{Amaro-Seoane}, P., {Andrews}, J., {Arca Sedda}, M., {et~al.} 2023, Living Reviews in Relativity, 26, 2, \dodoi{10.1007/s41114-022-00041-y}

\bibitem[{{Arcodia} {et~al.}(2021){Arcodia}, {Merloni}, {Nandra}, {Buchner}, {Salvato}, {Pasham}, {Remillard}, {Comparat}, {Lamer}, {Ponti}, {Malyali}, {Wolf}, {Arzoumanian}, {Bogensberger}, {Buckley}, {Gendreau}, {Gromadzki}, {Kara}, {Krumpe}, {Markwardt}, {Ramos-Ceja}, {Rau}, {Schramm}, \& {Schwope}}]{Arcodia+2021}
{Arcodia}, R., {Merloni}, A., {Nandra}, K., {et~al.} 2021, \nat, 592, 704, \dodoi{10.1038/s41586-021-03394-6}

\bibitem[{{Arcodia} {et~al.}(2024{\natexlab{a}}){Arcodia}, {Liu}, {Merloni}, {Malyali}, {Rau}, {Chakraborty}, {Goodwin}, {Buckley}, {Brink}, {Gromadzki}, {Arzoumanian}, {Buchner}, {Kara}, {Nandra}, {Ponti}, {Salvato}, {Anderson}, {Baldini}, {Grotova}, {Krumpe}, {Maitra}, {Miller-Jones}, \& {Ramos-Ceja}}]{Arcodia+2024a}
{Arcodia}, R., {Liu}, Z., {Merloni}, A., {et~al.} 2024{\natexlab{a}}, \aap, 684, A64, \dodoi{10.1051/0004-6361/202348881}

\bibitem[{{Arcodia} {et~al.}(2024{\natexlab{b}}){Arcodia}, {Merloni}, {Buchner}, {Baldini}, {Ponti}, {Rau}, {Liu}, {Nandra}, \& {Salvato}}]{Arcodia+2024b}
{Arcodia}, R., {Merloni}, A., {Buchner}, J., {et~al.} 2024{\natexlab{b}}, \aap, 684, L14, \dodoi{10.1051/0004-6361/202348949}

\bibitem[{{Arcodia} {et~al.}(2024{\natexlab{c}}){Arcodia}, {Linial}, {Miniutti}, {Franchini}, {Giustini}, {Bonetti}, {Sesana}, {Soria}, {Chakraborty}, {Dotti}, {Kara}, {Merloni}, {Ponti}, \& {Vincentelli}}]{Arcodia+2024c}
{Arcodia}, R., {Linial}, I., {Miniutti}, G., {et~al.} 2024{\natexlab{c}}, \aap, 690, A80, \dodoi{10.1051/0004-6361/202451218}

\bibitem[{{Arun} {et~al.}(2022){Arun}, {Belgacem}, {Benkel}, {Bernard}, {Berti}, {Bertone}, {Besancon}, {Blas}, {B{\"o}hmer}, {Brito}, {Calcagni}, {Cardenas-Avenda{\~n}o}, {Clough}, {Crisostomi}, {De Luca}, {Doneva}, {Escoffier}, {Ezquiaga}, {Ferreira}, {Fleury}, {Foffa}, {Franciolini}, {Frusciante}, {Garc{\'\i}a-Bellido}, {Herdeiro}, {Hertog}, {Hinderer}, {Jetzer}, {Lombriser}, {Maggio}, {Maggiore}, {Mancarella}, {Maselli}, {Nampalliwar}, {Nichols}, {Okounkova}, {Pani}, {Paschalidis}, {Raccanelli}, {Randall}, {Renaux-Petel}, {Riotto}, {Ruiz}, {Saffer}, {Sakellariadou}, {Saltas}, {Sathyaprakash}, {Shao}, {Sopuerta}, {Sotiriou}, {Stergioulas}, {Tamanini}, {Vernizzi}, {Witek}, {Wu}, {Yagi}, {Yazadjiev}, {Yunes}, {Zilh{\~a}o}, {Afshordi}, {Angonin}, {Baibhav}, {Barausse}, {Barreiro}, {Bartolo}, {Bellomo}, {Ben-Dayan}, {Bergshoeff}, {Bernuzzi}, {Bertacca}, {Bhagwat}, {Bonga}, {Burko}, {Comp{\'e}re}, {Cusin}, {da Silva}, {Das}, {de Rham}, {Destounis}, {Dimastrogiovanni}, {Duque}, {Easther}, {Farmer}, {Fasiello},
  {Fisenko}, {Fransen}, {Frauendiener}, {Gair}, {Gergely}, {Gerosa}, {Gualtieri}, {Han}, {Hees}, {Helfer}, {Hennig}, {Jenkins}, {Kajfasz}, {Kaloper}, {Karas}, {Kavanagh}, {Klioner}, {Koushiappas}, {Lagos}, {Poncin-Lafitte}, {Lobo}, {Markakis}, {Mart{\'\i}n-Moruno}, {Martins}, {Matarrese}, {Mayerson}, {Mimoso}, {Noller}, {Nunes}, {Oliveri}, {Orlando}, {Pappas}, {Pikovski}, {Pilo}, {Podolsk{\'y}}, {Pratten}, {Prokopec}, {Qi}, {Rastgoo}, {Ricciardone}, {Rollo}, {Rubiera-Garcia}, {Sergijenko}, {Shapiro}, {Shoemaker}, {Spallicci}, {Stashko}, {Stein}, {Tasinato}, {Tolley}, {Vagenas}, {Vandoren}, {Vernieri}, {Vicente}, {Wiseman}, {Zhdanov}, \& {Zumalac{\'a}rregui}}]{Arun+2022}
{Arun}, K.~G., {Belgacem}, E., {Benkel}, R., {et~al.} 2022, Living Reviews in Relativity, 25, 4, \dodoi{10.1007/s41114-022-00036-9}

\bibitem[{{Babak} {et~al.}(2017){Babak}, {Gair}, {Sesana}, {Barausse}, {Sopuerta}, {Berry}, {Berti}, {Amaro-Seoane}, {Petiteau}, \& {Klein}}]{Babak+2017}
{Babak}, S., {Gair}, J., {Sesana}, A., {et~al.} 2017, \prd, 95, 103012, \dodoi{10.1103/PhysRevD.95.103012}

\bibitem[{{Baker} {et~al.}(2022){Baker}, {Calcagni}, {Chen}, {Fasiello}, {Lombriser}, {Martinovic}, {Pieroni}, {Sakellariadou}, {Tasinato}, {Bertacca}, {Saltas}, \& {LISA Cosmology Working Group}}]{Baker+2022}
{Baker}, T., {Calcagni}, G., {Chen}, A., {et~al.} 2022, Journal of Cosmology and Astroparticle Physics, 2022, 031, \dodoi{10.1088/1475-7516/2022/08/031}

\bibitem[{{Berti} {et~al.}(2005){Berti}, {Buonanno}, \& {Will}}]{Berti+2005}
{Berti}, E., {Buonanno}, A., \& {Will}, C.~M. 2005, Classical and Quantum Gravity, 22, S943, \dodoi{10.1088/0264-9381/22/18/S08}

\bibitem[{{Bondi} \& {Hoyle}(1944)}]{Bondi&Hoyle1944}
{Bondi}, H., \& {Hoyle}, F. 1944, \mnras, 104, 273, \dodoi{10.1093/mnras/104.5.273}

\bibitem[{{Budnik} {et~al.}(2010){Budnik}, {Katz}, {Sagiv}, \& {Waxman}}]{Budnik+2010}
{Budnik}, R., {Katz}, B., {Sagiv}, A., \& {Waxman}, E. 2010, \apj, 725, 63, \dodoi{10.1088/0004-637X/725/1/63}

\bibitem[{{Bykov} {et~al.}(2024){Bykov}, {Gilfanov}, {Sunyaev}, \& {Medvedev}}]{Bykov+2024}
{Bykov}, S., {Gilfanov}, M., {Sunyaev}, R., \& {Medvedev}, P. 2024, arXiv e-prints, arXiv:2409.16908, \dodoi{10.48550/arXiv.2409.16908}

\bibitem[{{Cao}(2009)}]{Cao2009}
{Cao}, X. 2009, \mnras, 394, 207, \dodoi{10.1111/j.1365-2966.2008.14347.x}

\bibitem[{{Chakraborty} {et~al.}(2021){Chakraborty}, {Kara}, {Masterson}, {Giustini}, {Miniutti}, \& {Saxton}}]{Chakraborty+2021}
{Chakraborty}, J., {Kara}, E., {Masterson}, M., {et~al.} 2021, \apjl, 921, L40, \dodoi{10.3847/2041-8213/ac313b}

\bibitem[{{Chakraborty} {et~al.}(2024){Chakraborty}, {Arcodia}, {Kara}, {Miniutti}, {Giustini}, {Tetarenko}, {Rhodes}, {Franchini}, {Bonetti}, {Burdge}, {Goodwin}, {Maccarone}, {Merloni}, {Ponti}, {Remillard}, \& {Saxton}}]{Chakraborty+2024}
{Chakraborty}, J., {Arcodia}, R., {Kara}, E., {et~al.} 2024, \apj, 965, 12, \dodoi{10.3847/1538-4357/ad2941}

\bibitem[{{Chakraborty} {et~al.}(2025){Chakraborty}, {Kara}, {Arcodia}, {Buchner}, {Giustini}, {Hern\textbackslash'andez-Garc\textbackslash'ia}, {Linial}, {Masterson}, {Miniutti}, {Mummery}, {Panagiotou}, {Quintin}, \& {S\textbackslash'anchez-S\textbackslash'aez}}]{Chakraborty+2025}
{Chakraborty}, J., {Kara}, E., {Arcodia}, R., {et~al.} 2025, arXiv e-prints, arXiv:2503.19013, \dodoi{10.48550/arXiv.2503.19013}

\bibitem[{{Cutler}(1998)}]{Cutler1998}
{Cutler}, C. 1998, \prd, 57, 7089, \dodoi{10.1103/PhysRevD.57.7089}

\bibitem[{{Cutler} {et~al.}(2003){Cutler}, {Hiscock}, \& {Larson}}]{Cutler+2003}
{Cutler}, C., {Hiscock}, W.~A., \& {Larson}, S.~L. 2003, \prd, 67, 024015, \dodoi{10.1103/PhysRevD.67.024015}

\bibitem[{{Dai} {et~al.}(2010){Dai}, {Fuerst}, \& {Blandford}}]{Dai+2010}
{Dai}, L.~J., {Fuerst}, S.~V., \& {Blandford}, R. 2010, \mnras, 402, 1614, \dodoi{10.1111/j.1365-2966.2009.16038.x}

\bibitem[{{Damour} {et~al.}(2001){Damour}, {Iyer}, \& {Sathyaprakash}}]{Damour+2001}
{Damour}, T., {Iyer}, B.~R., \& {Sathyaprakash}, B.~S. 2001, \prd, 63, 044023, \dodoi{10.1103/PhysRevD.63.044023}

\bibitem[{{de Rham} \& {Melville}(2018)}]{deRham&Melville2018}
{de Rham}, C., \& {Melville}, S. 2018, \prl, 121, 221101, \dodoi{10.1103/PhysRevLett.121.221101}

\bibitem[{{Droz} {et~al.}(1999){Droz}, {Knapp}, {Poisson}, \& {Owen}}]{Droz+1999}
{Droz}, S., {Knapp}, D.~J., {Poisson}, E., \& {Owen}, B.~J. 1999, \prd, 59, 124016, \dodoi{10.1103/PhysRevD.59.124016}

\bibitem[{{Evans} \& {Kochanek}(1989)}]{Evans&Kochanek1989}
{Evans}, C.~R., \& {Kochanek}, C.~S. 1989, \apjl, 346, L13, \dodoi{10.1086/185567}

\bibitem[{{Fedderke} {et~al.}(2022){Fedderke}, {Graham}, \& {Rajendran}}]{Fedderke+2022}
{Fedderke}, M.~A., {Graham}, P.~W., \& {Rajendran}, S. 2022, \prd, 105, 103018, \dodoi{10.1103/PhysRevD.105.103018}

\bibitem[{{Foster} {et~al.}(2025){Foster}, {Blas}, {Bourgoin}, {Hees}, {Herrero-Valea}, {Jenkins}, \& {Xue}}]{Foster+2025}
{Foster}, J.~W., {Blas}, D., {Bourgoin}, A., {et~al.} 2025, arXiv e-prints, arXiv:2504.15334, \dodoi{10.48550/arXiv.2504.15334}

\bibitem[{{Franchini} {et~al.}(2023){Franchini}, {Bonetti}, {Lupi}, {Miniutti}, {Bortolas}, {Giustini}, {Dotti}, {Sesana}, {Arcodia}, \& {Ryu}}]{Franchini+2023}
{Franchini}, A., {Bonetti}, M., {Lupi}, A., {et~al.} 2023, \aap, 675, A100, \dodoi{10.1051/0004-6361/202346565}

\bibitem[{{Gaskin} {et~al.}(2019){Gaskin}, {Swartz}, {Vikhlinin}, {{\"O}zel}, {Gelmis}, {Arenberg}, {Bandler}, {Bautz}, {Civitani}, {Dominguez}, {Eckart}, {Falcone}, {Figueroa-Feliciano}, {Freeman}, {G{\"u}nther}, {Havey}, {Heilmann}, {Kilaru}, {Kraft}, {McCarley}, {McEntaffer}, {Pareschi}, {Purcell}, {Reid}, {Schattenburg}, {Schwartz}, {Schwartz}, {Tananbaum}, {Tremblay}, {Zhang}, \& {Zuhone}}]{Gaskin+2019}
{Gaskin}, J.~A., {Swartz}, D.~A., {Vikhlinin}, A., {et~al.} 2019, Journal of Astronomical Telescopes, Instruments, and Systems, 5, 021001, \dodoi{10.1117/1.JATIS.5.2.021001}

\bibitem[{{Generozov} \& {Perets}(2023)}]{Generozov&Perets2023}
{Generozov}, A., \& {Perets}, H.~B. 2023, \mnras, 522, 1763, \dodoi{10.1093/mnras/stad1016}

\bibitem[{{Gezari}(2021)}]{Gezari2021}
{Gezari}, S. 2021, \araa, 59, 21, \dodoi{10.1146/annurev-astro-111720-030029}

\bibitem[{{Giustini} {et~al.}(2020){Giustini}, {Miniutti}, \& {Saxton}}]{Giustini+2020}
{Giustini}, M., {Miniutti}, G., \& {Saxton}, R.~D. 2020, \aap, 636, L2, \dodoi{10.1051/0004-6361/202037610}

\bibitem[{{Guo} \& {Shen}(2025)}]{Guo&Shen2025}
{Guo}, W., \& {Shen}, R.-F. 2025, arXiv e-prints, arXiv:2504.12762, \dodoi{10.48550/arXiv.2504.12762}

\bibitem[{{Hawking} \& {Israel}(1987)}]{Hawking&Israel1987}
{Hawking}, S.~W., \& {Israel}, W. 1987, {Three hundred years of gravitation}

\bibitem[{{Hayasaki} \& {Yamazaki}(2019)}]{Hayasaki&Yamazaki2019}
{Hayasaki}, K., \& {Yamazaki}, R. 2019, \apj, 886, 114, \dodoi{10.3847/1538-4357/ab44ca}

\bibitem[{{Hern{\'a}ndez-Garc{\'\i}a} {et~al.}(2025){Hern{\'a}ndez-Garc{\'\i}a}, {Chakraborty}, {S{\'a}nchez-S{\'a}ez}, {Ricci}, {Cuadra}, {McKernan}, {Ford}, {Ar{\'e}valo}, {Rau}, {Arcodia}, {Kara}, {Liu}, {Merloni}, {Bruni}, {Goodwin}, {Arzoumanian}, {Assef}, {Baldini}, {Bayo}, {Bauer}, {Bernal}, {Brightman}, {Calistro Rivera}, {Gendreau}, {Homan}, {Krumpe}, {Lira}, {Mart{\'\i}nez-Aldama}, {Salvato}, \& {Sotomayor}}]{Hernandez-Garcia+2025}
{Hern{\'a}ndez-Garc{\'\i}a}, L., {Chakraborty}, J., {S{\'a}nchez-S{\'a}ez}, P., {et~al.} 2025, arXiv e-prints, arXiv:2504.07169, \dodoi{10.48550/arXiv.2504.07169}

\bibitem[{{Hoyle} \& {Lyttleton}(1939)}]{Hoyle&Lyttleton1939}
{Hoyle}, F., \& {Lyttleton}, R.~A. 1939, Proceedings of the Cambridge Philosophical Society, 35, 405, \dodoi{10.1017/S0305004100021150}

\bibitem[{{Ingram} {et~al.}(2021){Ingram}, {Motta}, {Aigrain}, \& {Karastergiou}}]{Ingram+2021}
{Ingram}, A., {Motta}, S.~E., {Aigrain}, S., \& {Karastergiou}, A. 2021, \mnras, 503, 1703, \dodoi{10.1093/mnras/stab609}

\bibitem[{{Ito} {et~al.}(2020){Ito}, {Levinson}, \& {Nagataki}}]{Ito+2020}
{Ito}, H., {Levinson}, A., \& {Nagataki}, S. 2020, \mnras, 492, 1902, \dodoi{10.1093/mnras/stz3591}

\bibitem[{{Ito} {et~al.}(2018){Ito}, {Levinson}, {Stern}, \& {Nagataki}}]{Ito+2018}
{Ito}, H., {Levinson}, A., {Stern}, B.~E., \& {Nagataki}, S. 2018, \mnras, 474, 2828, \dodoi{10.1093/mnras/stx2722}

\bibitem[{{Karas} \& {{\v{S}}ubr}(2001)}]{Karas&Subr2001}
{Karas}, V., \& {{\v{S}}ubr}, L. 2001, \aap, 376, 686, \dodoi{10.1051/0004-6361:20011009}

\bibitem[{{Kato} {et~al.}(2008){Kato}, {Fukue}, \& {Mineshige}}]{Kato+2008}
{Kato}, S., {Fukue}, J., \& {Mineshige}, S. 2008, {Black-Hole Accretion Disks --- Towards a New Paradigm ---}

\bibitem[{{Katz} {et~al.}(2010){Katz}, {Budnik}, \& {Waxman}}]{Katz+2010}
{Katz}, B., {Budnik}, R., \& {Waxman}, E. 2010, \apj, 716, 781, \dodoi{10.1088/0004-637X/716/1/781}

\bibitem[{{Kaur} {et~al.}(2023){Kaur}, {Stone}, \& {Gilbaum}}]{Kaur+2023}
{Kaur}, K., {Stone}, N.~C., \& {Gilbaum}, S. 2023, \mnras, 524, 1269, \dodoi{10.1093/mnras/stad1894}

\bibitem[{{Kejriwal} {et~al.}(2024){Kejriwal}, {Witzany}, {Zaja{\v{c}}ek}, {Pasham}, \& {Chua}}]{Kejriwal+2024}
{Kejriwal}, S., {Witzany}, V., {Zaja{\v{c}}ek}, M., {Pasham}, D.~R., \& {Chua}, A. J.~K. 2024, \mnras, 532, 2143, \dodoi{10.1093/mnras/stae1599}

\bibitem[{{Kim} \& {Kim}(2009)}]{Kim&Kim2009}
{Kim}, H., \& {Kim}, W.-T. 2009, \apj, 703, 1278, \dodoi{10.1088/0004-637X/703/2/1278}

\bibitem[{{King}(2020)}]{King2020}
{King}, A. 2020, \mnras, 493, L120, \dodoi{10.1093/mnrasl/slaa020}

\bibitem[{{Krolik} \& {Linial}(2022)}]{Krolik&Linial2022}
{Krolik}, J.~H., \& {Linial}, I. 2022, \apj, 941, 24, \dodoi{10.3847/1538-4357/ac9eb6}

\bibitem[{{Larson} \& {Hiscock}(2000)}]{Larson&Hiscock2000}
{Larson}, S.~L., \& {Hiscock}, W.~A. 2000, \prd, 61, 104008, \dodoi{10.1103/PhysRevD.61.104008}

\bibitem[{{Lee} {et~al.}(2014){Lee}, {Cunningham}, {McKee}, \& {Klein}}]{Lee+2014}
{Lee}, A.~T., {Cunningham}, A.~J., {McKee}, C.~F., \& {Klein}, R.~I. 2014, \apj, 783, 50, \dodoi{10.1088/0004-637X/783/1/50}

\bibitem[{{Levinson} \& {Nakar}(2020)}]{Levinson&Nakar2020}
{Levinson}, A., \& {Nakar}, E. 2020, \physrep, 866, 1, \dodoi{10.1016/j.physrep.2020.04.003}

\bibitem[{{Linial} \& {Metzger}(2023)}]{Linial&Metzger2023}
{Linial}, I., \& {Metzger}, B.~D. 2023, \apj, 957, 34, \dodoi{10.3847/1538-4357/acf65b}

\bibitem[{{Linial} \& {Metzger}(2024{\natexlab{a}})}]{Linial&Metzger2024b}
---. 2024{\natexlab{a}}, \apj, 973, 101, \dodoi{10.3847/1538-4357/ad639e}

\bibitem[{{Linial} \& {Metzger}(2024{\natexlab{b}})}]{Linial&Metzger2024a}
---. 2024{\natexlab{b}}, \apjl, 963, L1, \dodoi{10.3847/2041-8213/ad2464}

\bibitem[{{Linial} \& {Sari}(2023)}]{Linial&Sari2023}
{Linial}, I., \& {Sari}, R. 2023, \apj, 945, 86, \dodoi{10.3847/1538-4357/acbd3d}

\bibitem[{{Littenberg} \& {Cornish}(2019)}]{Littenberg&Cornish2019}
{Littenberg}, T.~B., \& {Cornish}, N.~J. 2019, \apjl, 881, L43, \dodoi{10.3847/2041-8213/ab385f}

\bibitem[{{Liu} {et~al.}(2003){Liu}, {Mineshige}, \& {Ohsuga}}]{Liu+2003}
{Liu}, B.~F., {Mineshige}, S., \& {Ohsuga}, K. 2003, \apj, 587, 571, \dodoi{10.1086/368282}

\bibitem[{{Liu} {et~al.}(2002){Liu}, {Mineshige}, \& {Shibata}}]{Liu+2002}
{Liu}, B.~F., {Mineshige}, S., \& {Shibata}, K. 2002, \apjl, 572, L173, \dodoi{10.1086/341877}

\bibitem[{{Lops} {et~al.}(2023){Lops}, {Izquierdo-Villalba}, {Colpi}, {Bonoli}, {Sesana}, \& {Mangiagli}}]{Lops+2023}
{Lops}, G., {Izquierdo-Villalba}, D., {Colpi}, M., {et~al.} 2023, \mnras, 519, 5962, \dodoi{10.1093/mnras/stad058}

\bibitem[{{Lu} \& {Quataert}(2023)}]{Lu&Quataert2023}
{Lu}, W., \& {Quataert}, E. 2023, \mnras, 524, 6247, \dodoi{10.1093/mnras/stad2203}

\bibitem[{{Lyu} {et~al.}(2024){Lyu}, {Pan}, {Mao}, {Jiang}, \& {Yang}}]{Lyu+2025}
{Lyu}, Z., {Pan}, Z., {Mao}, J., {Jiang}, N., \& {Yang}, H. 2024, arXiv e-prints, arXiv:2501.03252, \dodoi{10.48550/arXiv.2501.03252}

\bibitem[{{Maggiore}(2007)}]{Maggiore2007}
{Maggiore}, M. 2007, {Gravitational Waves: Volume 1: Theory and Experiments}, \dodoi{10.1093/acprof:oso/9780198570745.001.0001}

\bibitem[{{Metzger} {et~al.}(2022){Metzger}, {Stone}, \& {Gilbaum}}]{Metzger+2022}
{Metzger}, B.~D., {Stone}, N.~C., \& {Gilbaum}, S. 2022, \apj, 926, 101, \dodoi{10.3847/1538-4357/ac3ee1}

\bibitem[{{Middleton} {et~al.}(2025){Middleton}, {G{\'u}rpide}, {Kwan}, {Dai}, {Arcodia}, {Chakraborty}, {Dauser}, {Fragile}, {Ingram}, {Miniutti}, {Pinto}, \& {Kosec}}]{Middleton+2025}
{Middleton}, M., {G{\'u}rpide}, A., {Kwan}, T.~M., {et~al.} 2025, \mnras, 537, 1688, \dodoi{10.1093/mnras/staf052}

\bibitem[{{Miniutti} {et~al.}(2019){Miniutti}, {Saxton}, {Giustini}, {Alexander}, {Fender}, {Heywood}, {Monageng}, {Coriat}, {Tzioumis}, {Read}, {Knigge}, {Gandhi}, {Pretorius}, \& {Ag{\'\i}s-Gonz{\'a}lez}}]{Miniutti+2019}
{Miniutti}, G., {Saxton}, R.~D., {Giustini}, M., {et~al.} 2019, \nat, 573, 381, \dodoi{10.1038/s41586-019-1556-x}

\bibitem[{{Mushotzky} {et~al.}(2019){Mushotzky}, {Aird}, {Barger}, {Cappelluti}, {Chartas}, {Corrales}, {Eufrasio}, {Fabian}, {Falcone}, {Gallo}, {Gilli}, {Grant}, {Hardcastle}, {Hodges-Kluck}, {Kara}, {Koss}, {Li}, {Lisse}, {Loewenstein}, {Markevitch}, {Meyer}, {Miller}, {Mulchaey}, {Petre}, {Ptak}, {Reynolds}, {Russell}, {Safi-Harb}, {Smith}, {Snios}, {Tombesi}, {Valencic}, {Walker}, {Williams}, {Winter}, {Yamaguchi}, {Zhang}, {Arenberg}, {Brandt}, {Burrows}, {Georganopoulos}, {Miller}, {Norman}, \& {Rosati}}]{Mushotzky+2019}
{Mushotzky}, R., {Aird}, J., {Barger}, A.~J., {et~al.} 2019, in Bulletin of the American Astronomical Society, Vol.~51, 107, \dodoi{10.48550/arXiv.1903.04083}

\bibitem[{{Nakar} \& {Sari}(2010)}]{Nakar&Sari2010}
{Nakar}, E., \& {Sari}, R. 2010, \apj, 725, 904, \dodoi{10.1088/0004-637X/725/1/904}

\bibitem[{{Nandra} {et~al.}(2013){Nandra}, {Barret}, {Barcons}, {Fabian}, {den Herder}, {Piro}, {Watson}, {Adami}, {Aird}, {Afonso}, {Alexander}, {Argiroffi}, {Amati}, {Arnaud}, {Atteia}, {Audard}, {Badenes}, {Ballet}, {Ballo}, {Bamba}, {Bhardwaj}, {Stefano Battistelli}, {Becker}, {De Becker}, {Behar}, {Bianchi}, {Biffi}, {B{\^\i}rzan}, {Bocchino}, {Bogdanov}, {Boirin}, {Boller}, {Borgani}, {Borm}, {Bouch{\'e}}, {Bourdin}, {Bower}, {Braito}, {Branchini}, {Branduardi-Raymont}, {Bregman}, {Brenneman}, {Brightman}, {Br{\"u}ggen}, {Buchner}, {Bulbul}, {Brusa}, {Bursa}, {Caccianiga}, {Cackett}, {Campana}, {Cappelluti}, {Cappi}, {Carrera}, {Ceballos}, {Christensen}, {Chu}, {Churazov}, {Clerc}, {Corbel}, {Corral}, {Comastri}, {Costantini}, {Croston}, {Dadina}, {D'Ai}, {Decourchelle}, {Della Ceca}, {Dennerl}, {Dolag}, {Done}, {Dovciak}, {Drake}, {Eckert}, {Edge}, {Ettori}, {Ezoe}, {Feigelson}, {Fender}, {Feruglio}, {Finoguenov}, {Fiore}, {Galeazzi}, {Gallagher}, {Gandhi}, {Gaspari}, {Gastaldello}, {Georgakakis},
  {Georgantopoulos}, {Gilfanov}, {Gitti}, {Gladstone}, {Goosmann}, {Gosset}, {Grosso}, {Guedel}, {Guerrero}, {Haberl}, {Hardcastle}, {Heinz}, {Alonso Herrero}, {Herv{\'e}}, {Holmstrom}, {Iwasawa}, {Jonker}, {Kaastra}, {Kara}, {Karas}, {Kastner}, {King}, {Kosenko}, {Koutroumpa}, {Kraft}, {Kreykenbohm}, {Lallement}, {Lanzuisi}, {Lee}, {Lemoine-Goumard}, {Lobban}, {Lodato}, {Lovisari}, {Lotti}, {McCharthy}, {McNamara}, {Maggio}, {Maiolino}, {De Marco}, {de Martino}, {Mateos}, {Matt}, {Maughan}, {Mazzotta}, {Mendez}, {Merloni}, {Micela}, {Miceli}, {Mignani}, {Miller}, {Miniutti}, {Molendi}, {Montez}, {Moretti}, {Motch}, {Naz{\'e}}, {Nevalainen}, {Nicastro}, {Nulsen}, {Ohashi}, {O'Brien}, {Osborne}, {Oskinova}, {Pacaud}, {Paerels}, {Page}, {Papadakis}, {Pareschi}, {Petre}, {Petrucci}, {Piconcelli}, {Pillitteri}, {Pinto}, {de Plaa}, {Pointecouteau}, {Ponman}, {Ponti}, {Porquet}, {Pounds}, {Pratt}, {Predehl}, {Proga}, {Psaltis}, {Rafferty}, {Ramos-Ceja}, {Ranalli}, {Rasia}, {Rau}, {Rauw}, {Rea}, {Read}, {Reeves},
  {Reiprich}, {Renaud}, {Reynolds}, {Risaliti}, {Rodriguez}, {Rodriguez Hidalgo}, {Roncarelli}, {Rosario}, {Rossetti}, {Rozanska}, {Rovilos}, {Salvaterra}, {Salvato}, {Di Salvo}, {Sanders}, {Sanz-Forcada}, {Schawinski}, {Schaye}, {Schwope}, \& {Sciortino}}]{Nandra+2013}
{Nandra}, K., {Barret}, D., {Barcons}, X., {et~al.} 2013, arXiv e-prints, arXiv:1306.2307, \dodoi{10.48550/arXiv.1306.2307}

\bibitem[{{Narayan} \& {Yi}(1994)}]{Narayan&Yi1994}
{Narayan}, R., \& {Yi}, I. 1994, \apjl, 428, L13, \dodoi{10.1086/187381}

\bibitem[{{Narayan} \& {Yi}(1995)}]{Narayan&Yi1995}
---. 1995, \apj, 452, 710, \dodoi{10.1086/176343}

\bibitem[{{Nicholl} {et~al.}(2024){Nicholl}, {Pasham}, {Mummery}, {Guolo}, {Gendreau}, {Dewangan}, {Ferrara}, {Remillard}, {Bonnerot}, {Chakraborty}, {Hajela}, {Dhillon}, {Gillan}, {Greenwood}, {Huber}, {Janiuk}, {Salvesen}, {van Velzen}, {Aamer}, {Alexander}, {Angus}, {Arzoumanian}, {Auchettl}, {Berger}, {de Boer}, {Cendes}, {Chambers}, {Chen}, {Chornock}, {Fulton}, {Gao}, {Gillanders}, {Gomez}, {Gompertz}, {Fabian}, {Herman}, {Ingram}, {Kara}, {Laskar}, {Lawrence}, {Lin}, {Lowe}, {Magnier}, {Margutti}, {McGee}, {Minguez}, {Moore}, {Nathan}, {Oates}, {Patra}, {Ramsden}, {Ravi}, {Ridley}, {Sheng}, {Smartt}, {Smith}, {Srivastav}, {Stein}, {Stevance}, {Turner}, {Wainscoat}, {Weston}, {Wevers}, \& {Young}}]{Nicholl+2024}
{Nicholl}, M., {Pasham}, D.~R., {Mummery}, A., {et~al.} 2024, \nat, 634, 804, \dodoi{10.1038/s41586-024-08023-6}

\bibitem[{{Nomura} {et~al.}(2018){Nomura}, {Oka}, {Yamada}, {Takekawa}, {Ohsuga}, {Takahashi}, \& {Asahina}}]{Nomura+2018}
{Nomura}, M., {Oka}, T., {Yamada}, M., {et~al.} 2018, \apj, 859, 29, \dodoi{10.3847/1538-4357/aabe32}

\bibitem[{{Olejak} {et~al.}(2025){Olejak}, {Stegmann}, {de Mink}, {Valli}, {Sari}, \& {Justham}}]{Olejak+2025}
{Olejak}, A., {Stegmann}, J., {de Mink}, S.~E., {et~al.} 2025, arXiv e-prints, arXiv:2503.21995, \dodoi{10.48550/arXiv.2503.21995}

\bibitem[{{Ostriker}(1999)}]{Ostriker1999}
{Ostriker}, E.~C. 1999, \apj, 513, 252, \dodoi{10.1086/306858}

\bibitem[{{Pan} {et~al.}(2023){Pan}, {Li}, \& {Cao}}]{Pan+2023}
{Pan}, X., {Li}, S.-L., \& {Cao}, X. 2023, \apj, 952, 32, \dodoi{10.3847/1538-4357/acd180}

\bibitem[{{Pan} {et~al.}(2022){Pan}, {Li}, {Cao}, {Miniutti}, \& {Gu}}]{Pan+2022}
{Pan}, X., {Li}, S.-L., {Cao}, X., {Miniutti}, G., \& {Gu}, M. 2022, \apjl, 928, L18, \dodoi{10.3847/2041-8213/ac5faf}

\bibitem[{{Pan} \& {Yang}(2021)}]{Pan&Yang2021}
{Pan}, Z., \& {Yang}, H. 2021, \prd, 103, 103018, \dodoi{10.1103/PhysRevD.103.103018}

\bibitem[{{Phinney}(1989)}]{Phinney1989}
{Phinney}, E.~S. 1989, in IAU Symposium, Vol. 136, The Center of the Galaxy, ed. M.~{Morris}, 543

\bibitem[{{Qiao} \& {Liu}(2017)}]{Qiao&Liu2017}
{Qiao}, E., \& {Liu}, B.~F. 2017, \mnras, 467, 898, \dodoi{10.1093/mnras/stx121}

\bibitem[{{Qiao} \& {Liu}(2018)}]{Qiao&Liu2018}
---. 2018, \mnras, 477, 210, \dodoi{10.1093/mnras/sty652}

\bibitem[{{Quintin} {et~al.}(2023){Quintin}, {Webb}, {Guillot}, {Miniutti}, {Kammoun}, {Giustini}, {Arcodia}, {Soucail}, {Clerc}, {Amato}, \& {Markwardt}}]{Quintin+2023}
{Quintin}, E., {Webb}, N.~A., {Guillot}, S., {et~al.} 2023, \aap, 675, A152, \dodoi{10.1051/0004-6361/202346440}

\bibitem[{{Raj} \& {Nixon}(2021)}]{Raj&Nixon2021}
{Raj}, A., \& {Nixon}, C.~J. 2021, \apj, 909, 82, \dodoi{10.3847/1538-4357/abdc25}

\bibitem[{{Rees}(1988)}]{Rees1988}
{Rees}, M.~J. 1988, \nat, 333, 523, \dodoi{10.1038/333523a0}

\bibitem[{{Sesana} {et~al.}(2021){Sesana}, {Korsakova}, {Arca Sedda}, {Baibhav}, {Barausse}, {Barke}, {Berti}, {Bonetti}, {Capelo}, {Caprini}, {Garcia-Bellido}, {Haiman}, {Jani}, {Jennrich}, {Johansson}, {Khan}, {Korol}, {Lamberts}, {Lupi}, {Mangiagli}, {Mayer}, {Nardini}, {Pacucci}, {Petiteau}, {Raccanelli}, {Rajendran}, {Regan}, {Shao}, {Spallicci}, {Tamanini}, {Volonteri}, {Warburton}, {Wong}, \& {Zumalacarregui}}]{Sesana+2021}
{Sesana}, A., {Korsakova}, N., {Arca Sedda}, M., {et~al.} 2021, Experimental Astronomy, 51, 1333, \dodoi{10.1007/s10686-021-09709-9}

\bibitem[{{Shakura} \& {Sunyaev}(1973)}]{Shakura&Sunyaev1973}
{Shakura}, N.~I., \& {Sunyaev}, R.~A. 1973, \aap, 24, 337

\bibitem[{{Shen} \& {Matzner}(2014)}]{Shen&Matzner2014}
{Shen}, R.-F., \& {Matzner}, C.~D. 2014, \apj, 784, 87, \dodoi{10.1088/0004-637X/784/2/87}

\bibitem[{{Sicilia} {et~al.}(2022){Sicilia}, {Lapi}, {Boco}, {Spera}, {Di Carlo}, {Mapelli}, {Shankar}, {Alexander}, {Bressan}, \& {Danese}}]{Sicilia+2022}
{Sicilia}, A., {Lapi}, A., {Boco}, L., {et~al.} 2022, \apj, 924, 56, \dodoi{10.3847/1538-4357/ac34fb}

\bibitem[{{{\'S}niegowska} {et~al.}(2023){{\'S}niegowska}, {Grz{\c{e}}dzielski}, {Czerny}, \& {Janiuk}}]{Sniegowska+2023}
{{\'S}niegowska}, M., {Grz{\c{e}}dzielski}, M., {Czerny}, B., \& {Janiuk}, A. 2023, \aap, 672, A19, \dodoi{10.1051/0004-6361/202243828}

\bibitem[{{Spieksma} \& {Cannizzaro}(2025)}]{Spieksma&Cannizzaro2025}
{Spieksma}, T. F.~M., \& {Cannizzaro}, E. 2025, arXiv e-prints, arXiv:2504.08033, \dodoi{10.48550/arXiv.2504.08033}

\bibitem[{{Stone} \& {Metzger}(2016)}]{Stone&Metzger2016}
{Stone}, N.~C., \& {Metzger}, B.~D. 2016, \mnras, 455, 859, \dodoi{10.1093/mnras/stv2281}

\bibitem[{{Strubbe} \& {Quataert}(2009)}]{Strubbe&Quataert2009}
{Strubbe}, L.~E., \& {Quataert}, E. 2009, \mnras, 400, 2070, \dodoi{10.1111/j.1365-2966.2009.15599.x}

\bibitem[{{Suzuguchi} \& {Matsumoto}(in prep.)}]{Suzuguchi&Matsumoto2025}
{Suzuguchi}, T., \& {Matsumoto}, T. in prep.

\bibitem[{{Suzuguchi} {et~al.}(2024){Suzuguchi}, {Sugimura}, {Hosokawa}, \& {Matsumoto}}]{Suzuguchi+2024}
{Suzuguchi}, T., {Sugimura}, K., {Hosokawa}, T., \& {Matsumoto}, T. 2024, \apj, 966, 7, \dodoi{10.3847/1538-4357/ad34af}

\bibitem[{{Syer} \& {Ulmer}(1999)}]{Syer&Ulmer1999}
{Syer}, D., \& {Ulmer}, A. 1999, \mnras, 306, 35, \dodoi{10.1046/j.1365-8711.1999.02445.x}

\bibitem[{{Tagawa} \& {Haiman}(2023)}]{Tagawa&Haiman2023}
{Tagawa}, H., \& {Haiman}, Z. 2023, \mnras, 526, 69, \dodoi{10.1093/mnras/stad2616}

\bibitem[{{Takeda} {et~al.}(2019){Takeda}, {Nishizawa}, {Nagano}, {Michimura}, {Komori}, {Ando}, \& {Hayama}}]{Takeda+2019}
{Takeda}, H., {Nishizawa}, A., {Nagano}, K., {et~al.} 2019, \prd, 100, 042001, \dodoi{10.1103/PhysRevD.100.042001}

\bibitem[{{Thun} {et~al.}(2016){Thun}, {Kuiper}, {Schmidt}, \& {Kley}}]{Thun+2016}
{Thun}, D., {Kuiper}, R., {Schmidt}, F., \& {Kley}, W. 2016, \aap, 589, A10, \dodoi{10.1051/0004-6361/201527629}

\bibitem[{{Tsz-Lok Lam} {et~al.}(2025){Tsz-Lok Lam}, {Shibata}, {Kawaguchi}, \& {Pelle}}]{Tsz-Lok_Lam+2025}
{Tsz-Lok Lam}, A., {Shibata}, M., {Kawaguchi}, K., \& {Pelle}, J. 2025, arXiv e-prints, arXiv:2504.17016, \dodoi{10.48550/arXiv.2504.17016}

\bibitem[{{{\v{S}}ubr} \& {Karas}(1999)}]{Subr&Karas1999}
{{\v{S}}ubr}, L., \& {Karas}, V. 1999, aap, 352, 452, \dodoi{10.48550/arXiv.astro-ph/9910401}

\bibitem[{{Vurm} {et~al.}(2025){Vurm}, {Linial}, \& {Metzger}}]{Vurm+2025}
{Vurm}, I., {Linial}, I., \& {Metzger}, B.~D. 2025, \apj, 983, 40, \dodoi{10.3847/1538-4357/adb74d}

\bibitem[{{Wang} \& {Merritt}(2004)}]{Wang&Merritt2004}
{Wang}, J., \& {Merritt}, D. 2004, \apj, 600, 149, \dodoi{10.1086/379767}

\bibitem[{{Wevers} {et~al.}(2022){Wevers}, {Pasham}, {Jalan}, {Rakshit}, \& {Arcodia}}]{Wevers+2022}
{Wevers}, T., {Pasham}, D.~R., {Jalan}, P., {Rakshit}, S., \& {Arcodia}, R. 2022, \aap, 659, L2, \dodoi{10.1051/0004-6361/202243143}

\bibitem[{{Wevers} {et~al.}(2024){Wevers}, {French}, {Zabludoff}, {Fischer}, {Rowlands}, {Guolo}, {Dalla Barba}, {Arcodia}, {Berton}, {Bian}, {Linial}, {Miniutti}, \& {Pasham}}]{Wevers+2024}
{Wevers}, T., {French}, K.~D., {Zabludoff}, A.~I., {et~al.} 2024, \apjl, 970, L23, \dodoi{10.3847/2041-8213/ad5f1b}

\bibitem[{{Xian} {et~al.}(2021){Xian}, {Zhang}, {Dou}, {He}, \& {Shu}}]{Xian+2021}
{Xian}, J., {Zhang}, F., {Dou}, L., {He}, J., \& {Shu}, X. 2021, \apjl, 921, L32, \dodoi{10.3847/2041-8213/ac31aa}

\bibitem[{{Yao} {et~al.}(2025){Yao}, {Quataert}, {Jiang}, {Lu}, \& {White}}]{Yao+2025}
{Yao}, P.~Z., {Quataert}, E., {Jiang}, Y.-F., {Lu}, W., \& {White}, C.~J. 2025, \apj, 978, 91, \dodoi{10.3847/1538-4357/ad8911}

\bibitem[{{Zhou} {et~al.}(2024{\natexlab{a}}){Zhou}, {Huang}, {Guo}, {Li}, \& {Pan}}]{Zhou+2024a}
{Zhou}, C., {Huang}, L., {Guo}, K., {Li}, Y.-P., \& {Pan}, Z. 2024{\natexlab{a}}, \prd, 109, 103031, \dodoi{10.1103/PhysRevD.109.103031}

\bibitem[{{Zhou} {et~al.}(2025){Zhou}, {Pan}, \& {Jiang}}]{Zhou+Pan+Jiang2025}
{Zhou}, C., {Pan}, Z., \& {Jiang}, N. 2025, arXiv e-prints, arXiv:2504.11078, \dodoi{10.48550/arXiv.2504.11078}

\bibitem[{{Zhou} {et~al.}(2024{\natexlab{b}}){Zhou}, {Zeng}, \& {Pan}}]{Zhou+2024c}
{Zhou}, C., {Zeng}, Y., \& {Pan}, Z. 2024{\natexlab{b}}, arXiv e-prints, arXiv:2411.18046, \dodoi{10.48550/arXiv.2411.18046}

\bibitem[{{Zhou} {et~al.}(2024{\natexlab{c}}){Zhou}, {Zhong}, {Zeng}, {Huang}, \& {Pan}}]{Zhou+2024b}
{Zhou}, C., {Zhong}, B., {Zeng}, Y., {Huang}, L., \& {Pan}, Z. 2024{\natexlab{c}}, \prd, 110, 083019, \dodoi{10.1103/PhysRevD.110.083019}

\end{thebibliography}

\end{document}